\documentclass[traditabstract]{aa}
\usepackage{graphicx}
\usepackage{txfonts}
\begin{document}

\title{Invariant manifolds and the response of spiral arms in
barred galaxies}

\author{P. Tsoutsis  \inst{1,2} \and C. Kalapotharakos \inst{1}
\and C.Efthymiopoulos \inst{1} \and G. Contopoulos \inst{1}}

\institute{Research Center for Astronomy, Academy of Athens, Soranou
Efessiou 4,  115 27 Athens, Greece. \email{ptsoutsi@phys.uoa.gr,
ckalapot@phys.uoa.gr, cefthim@academyofathens.gr,
gcontop@academyofathens.gr} \and Department of Physics, University
of Athens, GR115 27 Athens, Greece}

\date{Received; accepted}

\abstract{The unstable invariant manifolds of the short-period
family of periodic orbits around the unstable Lagrangian points
$L_1$ and $L_2$ of a barred galaxy define loci in the configuration
space which take the form of a trailing spiral pattern. In previous
works we have explored the association of such a pattern to the
observed spiral pattern in $N$-body models of barred-spiral galaxies
and found it to be quite relevant. Our aims in the present paper
are: a) to investigate this association in the case of the
self-consistent models of Kaufmann \& Contopoulos (1996) which
provide  an approximation of real barred-spiral galaxies. b) to
examine the dynamical role played by each of the non-axisymmetric
components of the potential, i.e. the bar and the spiral
perturbation, and their consequences on the form of the invariant
manifolds, and c) to examine the relation of `response' models
of barred-spiral galaxies with the theory of the invariant
manifolds. Our method relies on calculating the invariant manifolds
for values of the Jacobi constant close to its value for $L_1$ and
$L_2$. Our main results are the following: a) The invariant
manifolds yield the correct form of the imposed spiral pattern
provided that their calculation is done with the spiral potential
term turned on. We provide a theoretical model explaining the form
of the invariant manifolds that supports the spiral structure. The
azimuthal displacement of the Lagrangian points with respect
to the bar's major axis is a crucial parameter in this modeling.
When this is taken into account, the manifolds necessarily
develop in a spiral-like domain of the configuration space,
delimited from below by the boundary of a banana-like non-permitted
domain, and from above either by rotational KAM tori or by cantori
forming a stickiness zone. On the contrary, if the whole
non-axisymmetric perturbation is artificially `aligned' with the bar
(i.e. there is no azimuthal shift of the Lagrangian manifolds), the
manifolds support a ring rather than a spiral structure. b) We
construct `spiral response' models on the basis of the theory of the
invariant manifolds and examine the connection of the latter to
the `response' models (Patsis 2006) used to fit real barred-spiral
galaxies, explaining how are the manifolds related to a number of
morphological features seen in such models.}


\titlerunning{Invariant manifolds and spiral arms}

\authorrunning{P. Tsoutsis et al}

\maketitle
\section{Introduction}
The ordered or chaotic nature of orbits in barred galaxies has been
the subject of many investigations in the literature (Contopoulos
1981; Pfenniger 1984; Sparke \& Sellwood 1987; Pfenniger \& Frendli
1991; Kaufmann \& Contopoulos 1996; Patsis et al. 1997; Fux 2001;
Pichardo et al. 2004; Kaufmann \& Patsis 2005). Interest to this
problem stems from the fact that the existence (and degree) of chaos
has direct consequences on the morphological features of a rotating
galaxy. In particular, the appearance of a large degree of chaos in
the corotation region is one of the main reasons for why the bars
terminate near corotation (Contopoulos 1981, see Contopoulos 2002,
pp. 473-474 for a review).

Beyond corotation, prominent structures such as rings or spiral
arms are commonly observed. The role of the chaotic orbits in the
dynamics of such structures is still a widely open problem, but
recently some progress was made towards its understanding. In
particular, a theoretical model has been proposed and numerically
explored (Voglis et al. 2006a, 2006b; Romero-Gomez et al. 2006,
2007), according to which the spiral arms (or rings) are supported
by the {\it unstable invariant manifolds} of the two short period
families of unstable periodic orbits around the unstable
Lagrangian equilibria $L_1$ and $L_2$ (called hereafter the PL1
and PL2 families respectively). This theory was extended by
Tsoutsis et al. (2008), by examining the contribution of the
unstable manifolds of other families, besides PL1 or PL2, to the
same phenomenon. The importance of the chaotic orbits in
supporting the spiral structure of barred galaxies has also been
emphasized by Patsis (2006).

The following is a brief account of the theory of the invariant
manifolds:

1) We consider a 2D approximation of the orbits in the disk plane of
a barred - spiral galaxy, given by the Hamiltonian
\begin{equation}\label{hamgen}
H(r,\theta,p_r,p_{\theta})\equiv{1\over
2}\big(p_r^2+{p_\theta^2\over r^2}\big) -\Omega_p p_\theta
+V_0(r)+V_1(r,\theta)=E_J~~.
\end{equation}
In this expression, $(r,\theta)$ are polar coordinates in the
rotating frame, $p_r=\dot{r}$, $p_\theta=r^2(\dot{\theta}+\Omega_p)$
is the angular momentum in the rest frame, $V_0$ is the axisymmetric
potential and $V_1$ is the non-axisymmetric potential perturbation
due to the bar and to the spiral arms. $\Omega_p$ is the angular
speed of the rotating frame, which coincides with the bar-spiral
pattern speed in an approximation in which the latter is assumed to
be unique.

2) The Hamiltonian flow under (\ref{hamgen}) yields two stable
($L_4$, $L_5$) and two unstable ($L_1$, $L_2$) Lagrangian
equilibrium points (in the rotating frame) at which a star corotates
with the pattern. The {\it unstable manifold} ${\cal W}^U_{L_1}$ of
$L_1$ is defined as the set of all the initial conditions
$(r_0,\theta_0,p_{r0},p_{\theta0})$ in the phase space for which the
resulting orbit tends asymptotically to $L_1$ in the backward sense
of time, namely
\begin{eqnarray}\label{wul1}
{\cal W}^U_{L_1} &=
&\big\{\bigcup(r_0,\theta_0,p_{r0},p_{\theta0}):\nonumber\\
& &\lim_{t\rightarrow -\infty}||Q(t;r_0,\theta_0,p_{r0},p_{\theta0})
-(r_{L_1},\theta_{L_1},0,\Omega_p)||=0 \big\}
\end{eqnarray}
where $Q(t;r_0,\theta_0,p_{r0},p_{\theta0})$ denotes the position
(point in phase space) at time $t$ of a particle along an orbit
starting with the above initial conditions, and the norm $||\cdot||$
means the Euclidean distance between this point and the phase space
point $(r_{L_1},\theta_{L_1},0,\Omega_p)$, corresponding to $L_1$.
All the points of the manifold ${\cal W}^U_{L_1}$ yield the same
value of the Jacobi constant, equal to $E_J=E_{J,L1}$. Furthermore,
since $L_1$ is simply unstable, ${\cal W}^U_{L_1}$ is a
two-dimensional manifold embedded in the three-dimensional
hypersurface of the phase space corresponding to a fixed Jacobi
constant $E_J=E_{J,L1}$. Similar definitions and properties hold for
$L_2$ and ${\cal W}^U_{L_2}$ and for the stable manifolds ${\cal
W}^S_{L_1}$, ${\cal W}^S_{L_2}$, i.e. the sets of initial conditions
tending asymptotically to $L_1$, or $L_2$ in the forward sense of
time, as $t\rightarrow\infty$.

\begin{figure*}
\centering
\includegraphics[scale=1]{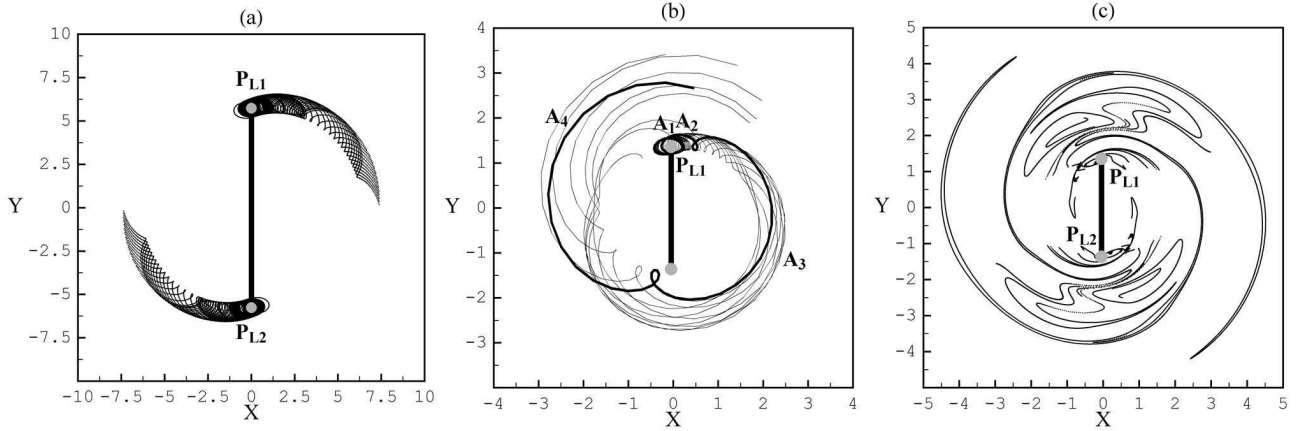}
\caption{ (a) Projections of the invariant manifolds ${\cal
W}^U_{PL1}$ and ${\cal W}^U_{PL2}$ on the configuration space,
approximated by an ensemble of 20 orbits with initial conditions on
the unstable direction of the tangent space to the periodic orbits
PL1 and PL2 (small bold circles) in the neighborhood of $L_1$ and
$L_2$. The potential used is from model A (see Sect. 2). The bar is
aligned to the y-axis and rotates counterclockwise. (b) Same as in
(a) but in a strongly non-linear model (in this case we use the
potential of the $N$-Body simulation analyzed in Voglis et al.
2006a. Only the manifold ${\cal W}^U_{PL1}$ is plotted). (c) The
projection of the intersection of the manifolds  ${\cal W}^U_{PL1}$
and ${\cal W}^U_{PL2}$ with the apocentric surface of section
$p_r=0$, $\dot{p}_r<0$ on the configuration space of the model of
(b).} \label{}
\end{figure*}
3) For $E_J>E_{J,L1}$, a short-period unstable periodic orbit
(PL1) bifurcates from $L_1$ (and the symmetric orbit PL2 from
$L_2$). This orbit forms a small loop around $L_1$ (Fig.1a, thick
solid curve), which corresponds to a 1D-torus in the phase space.
This torus is `whiskered', i.e., it possesses its own asymptotic
manifolds. In particular, the unstable manifold of PL1 is now
defined as
\begin{eqnarray}\label{wu}
{\cal W}^U_{PL1} &= &
\big\{\bigcup(r_0,\theta_0,p_{r0},p_{\theta0}):\nonumber\\
& &\lim_{t\rightarrow
-\infty}||Q(t;r_0,\theta_0,p_{r0},p_{\theta0})-PL1||=0 \big\}
\end{eqnarray}
where the notation $||\cdot||$ refers to the minimum of the
distances of $Q(t)$ from the locus of all the phase space points of
the orbit PL1. For any fixed value of $E_J>E_{J,L1}$, ${\cal
W}^U_{PL1}$ is a two-dimensional manifold embedded in the
three-dimensional hypersurface of constant $E_J$. Figure 1a shows
the projection of  a small part of this manifold, close to PL1, in
the {\it configuration space} $x=r\cos\theta$, $y=r\sin\theta$. This
is drawn approximately, by calculating a number of orbits with
initial conditions on ${\cal W}^U_{PL1}$, and close to PL1. The
possibility to find such initial conditions is guaranteed by the
fact that the manifold ${\cal W}^U_{PL1}$ is tangent to the unstable
manifold of the linearized Hamiltonian flow near PL1 (the so-called
Grobman 1959 and Hartman 1960 theorem), and the latter is calculated
by diagonalizing the Floquet matrix of the orbit PL1. In Fig.1a we
draw the part of the manifold lying outside corotation for a
particular model of barred galaxy.  We can see that close to PL1 the
orbits form epicyclic loops of size nearly equal to the PL1 loop,
while, in the same time, the guiding center recedes from PL1 along a
path which yields a {\it trailing spiral arm}. An analysis of the
linearized flow yields that the deviation of the guiding center from
PL1 is exponential in time, with a rate determined by the positive
characteristic exponents of the Floquet matrix of PL1. Furthermore,
in generic galactic potentials all the orbits on ${\cal W}^U_{PL1}$
are {\it chaotic}. (The same phenomena hold for the orbit PL2 and
the manifold ${\cal W}^U_{PL2}$ also plotted in Fig.1a)

4) In strongly nonlinear models (as is the case of strongly barred
galaxies with conspicuous spiral arms), further away from PL1 the
size of the epicycles becomes great (it may exceed the size of the
bar). Such an example is shown in Fig.1b, referring to the orbits of
the ${\cal W}^U_{PL1}$ family in a $N$-Body model of a barred galaxy
(Voglis et al. 2006a, 2006b). We see that one such orbit (bold)
forms two relatively small loops near PL1, reaching the apocentric
positions $A_1$ and $A_2$, but the exponential recession of the
guiding center is so fast that there is no loop formed between the
second and third ($A_3$) apocentric positions. Furthermore, the
fourth apocentric position is at a distance about twice the bar's
major semi-axis. Further integration beyond that of Fig.1b shows
that, in fact, all these orbits belong to the so-called `hot
population' (Sparke \& Sellwood 1987), i.e., the orbits make several
consecutive oscillations in and out of corotation. Kaufmann and
Contopoulos (1996, their figure 21a) suggested that such orbits can
partly support the bar and partly the spiral arms.

To understand how the chaotic orbits may establish a long-time flow
supporting the spiral structure, Voglis et al. (2006a) examined a
particular subset of points of the unstable manifold ${\cal
W}^U_{PL1}$, namely the locus of all {\it apocentric positions} of
the orbits on ${\cal W}^U_{PL1}$. This is defined by taking the
intersection of ${\cal W}^U_{PL1}$ with the so-called {\it surface
of section} of the apocentric positions on a hypersurface of
constant $E_J$, defined as the set of phase space points satisfying
$p_r=0$, $\dot{p}_r<0$. We stress that the choice of such a surface
of section proves to be very relevant to the particular type of
study undertaken here. Indeed, whenever $p_r=0$, Eq.(\ref{hamgen})
yields points $(r,\theta,p_\theta)$ which can be projected either on
the $(\theta,p_\theta)$ plane (called the `phase portrait' by virtue
of the fact that the variables $(\theta,p_\theta)$ are canonically
conjugated), or the plane $(\theta,r)$, i.e. the usual plane of
motion in the rotating frame. However, in typical galactic
potentials the variables $r$ and $p_\theta$ at the apocenters have a
monotonic relation, and this implies that the phase portraits and
$(r,\theta)$ portraits for this particular type of surface of
section are isomorphic. This allows us to unravel immediately the
consequences of the phase space dynamical features, as seen in the
phase portraits, to the morphological features of the system, as
seen in the usual disk plane of motion with coordinates
$(x,y)=(r\cos\theta,r\sin\theta)$. Such a comparison would not be
possible by the use of a traditional surface of section, such as
$(x,\dot{x})$ for $y=0,\dot{y}>0$ (or $\dot{y}<0$).

Returning to the role of the invariant manifolds, the intersection
of ${\cal W}^U_{PL1}$ with the apocentric surface of section yields
an one-dimensional locus of points. Such a locus can be projected on
either the phase portrait plane $(\theta,p_\theta)$ or the
configuration space $(r,\theta)$. Figure 1c shows the latter
projection in the case of the same manifold as in Fig.1b, but
calculated for a much larger length. Every point in Fig.1c
corresponds to one apocentric position of a chaotic orbit with
initial conditions on the unstable manifold. Clearly, the apocentric
positions along ${\cal W}^U_{PL1}$ yield a locus which also supports
a trailing spiral arm over, however, a much larger extent of ${\cal
W}^U_{PL1}$ than in the case of Fig.1b. The manifold of Fig.1c takes
a typical form known in dynamical systems' theory to be associated
with the so-called phenomenon of {\it homoclinic chaos}. Briefly,
the manifold develops lobes forming oscillations close to the
apocentric points of the periodic orbits PL1 or PL2. Such
oscillations are analyzed in detail in the sequel.

We should stress that an analysis of the Floquet matrix of the PL1
or PL2 families yields that only the directions of the unstable
invariant manifolds ${\cal W}^U_{PL1}$, ${\cal W}^U_{PL2}$ are such
as to define trailing spiral arms, while, close to $L_1$ or $L_2$,
the stable manifolds ${\cal W}^S_{PL1}$, ${\cal W}^S_{PL2}$ define
leading spiral arms. Furthermore, in the forward sense of time the
chaotic orbits are attracted in directions of the phase space along
the unstable manifolds. In the sequel we no longer refer to the
stable manifolds ${\cal W}^S_{PL1}$, ${\cal W}^S_{PL2}$, and the
term `invariant manifolds' always implies the unstable manifolds
${\cal W}^U_{PL1}$, ${\cal W}^U_{PL2}$ .

In summary, the theory of the invariant manifolds, viewed as either
the loci on which lies the continuous flow of a swarm of orbits
(Romero-Gomez et al. 2006, 2007), or the loci of apocentric
positions of these orbits (Voglis et al. 2006a; Tsoutsis et al.
2008), predicts the formation by the manifolds of a trailing spiral
pattern beyond corotation. Naturally, the central question that
should be posed now is whether (and up to what extent) the spiral
arms formed self-consistently in real galaxies can be associated
with the spiral patterns formed by the invariant manifolds ${\cal
W}^U_{PL1,2}$. In our previous works (Voglis et al. 2006a; Tsoutsis
et al. 2008), we examined this question by considering the spiral
arms formed in an $N$-Body model of a barred galaxy and found such
an association to be quite relevant.

In the present paper, our main goal is to examine the same question
in simple models of real barred-spiral galaxies for which some
reliable estimation of both the gravitational potential and the
pattern speed have been provided in the literature by methods
independent of the previous considerations. To this end, we selected
the potential models and pattern speeds reported in the study of
Kaufmann \& Contopoulos (1996) for three real galaxies, NGC3992,
NGC1073 and NGC1398. This choice is motivated by the fact that
Kaufmann \& Contopoulos (1996) constructed approximate
self-consistent models of the studied galaxies based on the response
density of the superposition of many stellar dynamical orbits. Thus,
their study yielded not only plausible values of the potential
parameters, or the pattern speed, but also the decomposition of the
potential into components, i.e., $V_{halo}$, $V_{disk}$, $V_{bar}$
and $V_{spiral}$. This allows us to check the role of each of these
components, in particular of the non-axisymmetric ones $V_{bar}$ and
$V_{spiral}$, in the theory. It should be noted that the
self-consistent technique, pioneered by  Schwarzschild (1979), has
been used extensively to provide reliable models of galaxies,
despite the fact that there is no a priori guarantee of the
stability of such models that should ideally be probed via $N$-body
simulations (see e.g. Smith and Miller 1982).

Besides re-confirming that the invariant manifolds do correlate well
with the spiral arms found in the self-consistent models of Kaufmann
\& Contopoulos (1996), our investigation led to a second non-trivial
result analyzed in detail in the sequel: In all three models the bar
component is dominant over the spiral component within a large
radial extent, but not in a narrow zone beyond corotation. This
implies that if one uses only the bar component to calculate the
manifolds, the latter yield ring rather than spiral structures.
Furthermore, if one adds the spiral perturbation to the potential,
but gives no azimuthal tilting to the associated $m=2$ Fourier
component, the manifolds become more open as regards their radial
extent, but remain quite symmetric as regards their orientation with
respect to the bar's major axis, thus still defining rings rather
than spiral arms. Only when the azimuthal deformation of the
equipotential surfaces due to a really spiral-like perturbation is
taken into account (in the Kaufmann \& Contopoulos (1996) paper this
was modeled as a simple logarithmic spiral), the manifolds are found
to follow closely the spiral arms of the self-consistent models. In
some numerical experiments (see Sect. 3 below) we managed to obtain
a kind of spiral pattern formed by the initial segments of the
invariant manifolds in pure bar models, having, however, to
drastically depart from the bar parameters given in Kaufmann and
Contopoulos' self-consistent models, and pushing the bar's amplitude
to highly non-physical values. But even in that case, the
manifold-induced spiral arms are quite different from the spiral
arms of the self-consistent models, and they disappear when the
manifolds are computed for a longer length. Such an investigation
demonstrates that while in principle the strength of the quadrupole
moment of the bar's potential causes a `thickening' of ring
structures, thus facilitating the phenomenon of appearance of spiral
arms (Romero-Gomez et al. 2007), this parameter is not sufficient in
order to characterize this phenomenon. The azimuthal
displacement of the Lagrangian points is the most important
parameter. This result probably provides a dynamical basis for
understanding the reported failure of pure bar models to reproduce
the inner spiral arms emanating at the ends of bars in both particle
and hydrodynamical simulations of barred galaxies (e.g. Lindblad et
al. 1996; Aguerri et al. 2001).

The paper is organized as follows: Section 2 gives the form of the
invariant manifolds in the models of Kaufmann and Contopoulos.
We examine the manifolds a) when the spiral perturbation is
turned-on, and b) in an `aligned model' version in which the whole
non-axisymmetric perturbation is artificially aligned to the bar. In
case (a) the manifolds yield a spiral response, while in case (b)
they yield a ring-like response. Since in all the above models the
spiral perturbation is strong, we also examine two models
corresponding to a `mean' and `weak' spiral amplitude, created by
suitably varying the parameters of some of the original models of
Kaufmann and Contopoulos (1996). We finally provide a theoretical
justification of the importance of the azimuthal displacement of the
Lagrangian points in the form of the invariant manifolds. Section 3
discusses the connection between the theory of the invariant
manifolds and the `spiral response' models constructed via iterative
methods. In particular, we propose a method of constructing response
models on the basis of populating by matter the manifolds generated
by a `pure bar' model. We also calculate response models via the
method proposed by Patsis (2006) and discuss a number of
morphological features of these models which find a straightforward
explanation by the invariant manifolds. Section 4 summarizes our
conclusions.

\section{Model and invariant manifolds}

\subsection{Model}
The model of Kaufmann \& Contopoulos (1996) consists of a number of
potential/density terms representing various components of a
barred-spiral galaxy. In particular we have:

- A halo density term given by a Plummer sphere
\begin{equation}\label{halodel}
\rho_h(r)={3M_h\over 4\pi b_h^3}\Bigg(1+{r^2\over
b_h^2}\Bigg)^{-5/2}~~.
\end{equation}

- A disk surface density given by an exponential law
\begin{equation}\label{diskden}
\Sigma_d(r)=\Sigma_0\exp(-\epsilon_d r)~~.
\end{equation}

- A Ferrers bar with major axis aligned with the y-axis
\begin{equation}\label{barden}
\rho_b(x,y,z)={105M_b\over 32\pi abc}\Bigg(1-{y^2\over
a^2}-{x^2\over b^2}-{z^2\over c^2}\Bigg)^2, ~~~~~a>b>c,
\end{equation}
with
$$
1-{y^2\over a^2}-{x^2\over b^2}-{z^2\over c^2}\geq 0
$$

 - and a spiral perturbation in the potential
\begin{equation}\label{spirpot}
V_s(r,\theta)=A(r)r\exp(-\epsilon_s r)\cos 2\Phi
\end{equation}
where
\begin{eqnarray}\label{phase}
\Phi &= &{\ln(r/a)\over \tan i}-\theta ~~~~~\mbox{if
$r\geq a$} \nonumber\\
\Phi &= &\theta ~~~~~~~~~~~~~~~~~~~~\mbox{if $r<a$}
\end{eqnarray}
and
\begin{equation}\label{aspir}
A(r)=\bigg({A-A_r\over 4}\bigg)
\big(1+\tanh[\kappa_1(r-r_1)]\big)\big(1+\tanh[\kappa_2(r_2-r)]\big)+A_r.
\end{equation}
The latter formula allows for a nearly constant amplitude $A(r)$
between the inner and outer cut-off radii $r_1$ and $r_2$, while the
amplitude falls to a small value $A_r$ beyond $r_2$ or below $r_1$.
We have in fact slightly modified Eq.(\ref{phase})
so that the potential becomes a smooth function of $\theta$ at $r=a$.
Namely, we substitute (\ref{phase}) by the expression:
\begin{equation}\label{phasen}
\Phi = {1\over 2}\bigg[1+\tanh(2(r-a))\bigg]{\ln(r/a)\over \tan
i}-\theta~~.
\end{equation}
The latter expression introduces a smoothing of the potential
yielding a difference with respect to Eq.(\ref{phase}) which is 10\%
at the distance $r=a\pm 0.5$Kpc and only 2\% at $r=a\pm 1$Kpc.

\begin{table}[h]
\caption{Parameters of models A, B, C (from Kaufmann \& Contopoulos
1996). The units are $\mbox{km}^2\mbox{s}^{-2}\mbox{kpc}^{-1}$ for
$A$, $\mbox{kpc}^{-1}$ for $\varepsilon_s$, $\varepsilon_d$,
$\kappa_1$ and $\kappa_2$, $\mbox{kpc}$ for $r_1$, $r_2$, $\Delta$,
$a$, $b$, $c$, and $b_h$, $10^{10}M_\odot$ for $M_h$, $M_b$,
$\mbox{km}\mbox{s}^{-1}\mbox{kpc}^{-1}$ for $\Omega_p$, and
$M_\odot/pc^2$ for $\Sigma_0$. Model A$'$ has the same
parameters as model A, except for the spiral amplitude $A=1000$ and
the pitch angle $i_0=-9^\circ$. Model B$'$ has the same parameters
as model B except for $A=2500$, $M_b=0.09$, $i_0=-8^\circ$,
$\Omega_p=30.5~$$\mbox{km}\mbox{s}^{-1}\mbox{kpc}^{-1}$.}
\begin{tabular}[c]{c c c c c c c c c}
  \hline \hline
  \multicolumn{9}{c}{Model A} \\
  \hline \hline
  Spiral: & A & $\varepsilon_s$ & $i_0$ & $r_1$ & $r_2$ & $\kappa_1$ & $\kappa_2$ & $\Delta$ \\
    & 2000 & 0.4 & -10$^\circ$ & 1.5 & 10.6 & 1 & 1 & 0.1 \\
  Bar: & $M_b$ & $a$ & $b$ & $c$ & $\Omega_p$ &  &   &   \\
    & 1.5 & 5.5 & 2.1 & 0.55 & 43.6 &   &   &   \\
  Disk: & $\Sigma_0$ & $\varepsilon_d$ &   &   &   &   &   &   \\
    &  750 & 0.235 &   &   &   &   &   &   \\
  Halo: & $M_h$ & $b_h$ &   &   &   &   &   &   \\
   & 27.5 & 12 &   &   &   &   &   &   \\
  \hline \hline
  \multicolumn{9}{c}{Model B} \\
  \hline \hline
  Spiral: & A & $\varepsilon_s$ & $i_0$ & $r_1$ & $r_2$ & $\kappa_1$ & $\kappa_2$ & $\Delta$ \\
    & 9000 & 1.2 & -10$^\circ$ & 3 & 5.6 & 0.5 & 0.5 & 0.1 \\
  Bar: & $M_b$ & $a$ & $b$ & $c$ & $\Omega_p$ &  &   &   \\
    & 0.18 & 3 & 0.4 & 0.3 & 32.2 &   &   &   \\
  Disk: & $\Sigma_0$ & $\varepsilon_d$ &   &   &   &   &   &   \\
    &  250 & 0.305 &   &   &   &   &   &   \\
  Halo: & $M_h$ & $b_h$ &   &   &   &   &   &   \\
   & 1.0 & 9 &   &   &   &   &   &   \\
  \hline \hline
    \multicolumn{9}{c}{Model C} \\
  \hline \hline
  Spiral: & A & $\varepsilon_s$ & $i_0$ & $r_1$ & $r_2$ & $\kappa_1$ & $\kappa_2$ & $\Delta$ \\
    & 3000 & 0.4 & -6$^\circ$ & 1 & 10 & 0.5 & 0.5 & 0.1 \\
  Bar: & $M_b$ & $a$ & $b$ & $c$ & $\Omega_p$ &  &   &   \\
    & 1.2 & 4.8 & 1.6 & 0.48 & 54.2 &   &   &   \\
  Disk: & $\Sigma_0$ & $\varepsilon_d$ &   &   &   &   &   &   \\
    &  1475 & 0.185 &   &   &   &   &   &   \\
  Halo: & $M_h$ & $b_h$ &   &   &   &   &   &   \\
   & 60.0 & 35 &   &   &   &   &   &   \\
  \hline
\end{tabular}
\end{table}
A model is specified by a set of values for the parameters $M_h$,
$b_h$, $\Sigma_0$, $\epsilon_d$, $M_b$, $a, b, c$, $\epsilon_s$,
$i$, $A$, $A_r$, $\kappa_1$, $r_1$, $\kappa_2$, and $r_2$, as well
as the value of the pattern angular speed $\Omega_p$. In Kaufmann \&
Contopoulos (1996), the parameters were adjusted so as to
produce three different self-consistent models which present some
features of three real barred galaxies. The criterion for
self-consistency was that the `response density', i.e., the density
obtained by the superposition of many orbits in the fixed potential
should match as closely as possible the imposed density represented
by the above equations. The matching refers to a) the amplitudes of
the surface density map on the disk plane, and b) the phases of the
maxima of the bar and of the spiral arms, in the imposed and in the
response models.

\begin{figure*}
\centering
\includegraphics[scale=1]{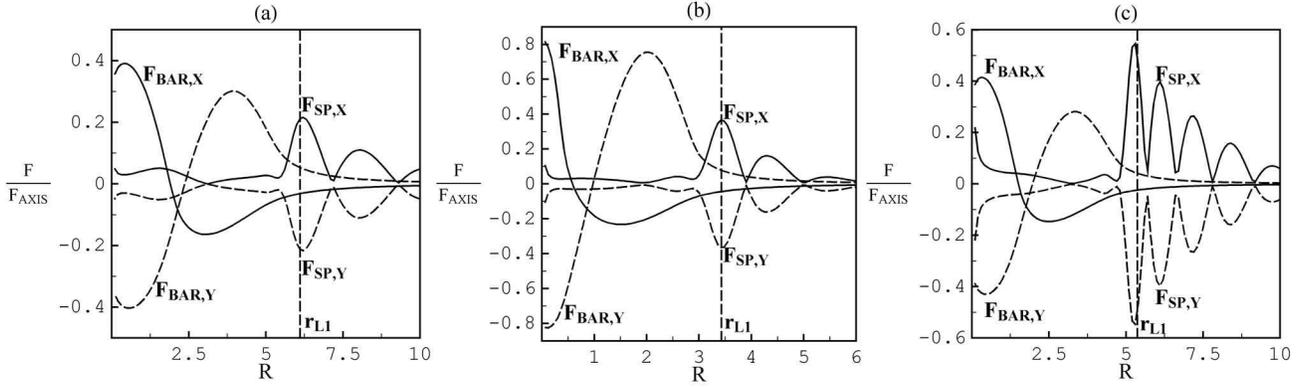}
\caption{The ratio of non-axisymmetric forces, due to the bar or
to the spiral arms, versus total axisymmetric force, as a function
of the distance $R$ along the x-axis (solid) or y-axis (dashed)
for models A, B, and C (panels a, b, and c respectively).
The vertical dashed lines mark the distance of the $L_1$ or $L_2$
points in each case.} \label{}
\end{figure*}

The parameters for the three models are given in Table 1. In
the sequel we refer to these as model A, B, and C. The value of
$A_r$ for all three models, as well as the values of $r_2$ and
$\kappa_2$ for model C are missing from Kaufmann \& Contopoulos
(1996), where, however, it is noted that any (small) value of $A_r$,
or of $r_1, r_2$ and $\kappa_1, \kappa_2$ does not influence the
self-consistency (provided that $r_2$ is at the end of the spiral
arms). For consistency with the remaining models, we have set
$\kappa_2=\kappa_1$ and $r_2=10$~Kpc in the case of model C, and
$A_r=0$ in all three models.

Models A, B, C present some features of the galaxies NGC3992,
NGC1073, and NGC1398 respectively. As discussed below, the
bar-spiral strengths induced by the parameters of model C are quite
untypical of barred-spiral galaxies, although still in the range
allowed by observations. Thus, while the theory of the invariant
manifolds worked well in all three models, we discuss in detail
models A and B, and only some exceptional features of model C
interesting for dynamics. It should be pointed out that, while the
choice of model parameters was partly based on observations (see
Sects. 2, 3 of Kaufmann \& Contopoulos 1996), the so-obtained models
are only rough representations of the referenced galaxies. For
example, images of the galaxy NGC3992 (e.g. in the I band, Tully et.
al. 1996) indicate the presence of at least one more arm of
amplitude comparable to the main bi-symmetric pattern. Images of the
galaxy NGC1398, (e.g. in the R-band, Hammed \& Devereux 1999) reveal
the existence of an inner ring structure which has no clear-cut
separation from the main spiral structure. Such morphological
features are not captured by the potential/density model given by
Eqs. (\ref{halodel})--(\ref{phasen}). Finally, the use of a $n=2$
Ferrers bar model implies a steep drop of the bar force beyond the
bar's limit which would be smoother in a $n=0$ or $n=1$ model,
and it does also not account for a rectangular-like outline that is
observed in many real bars. 

These facts notwithstanding, the
choice of potential parameters and pattern speeds as in Table 1
ensures the existence of a self-consistent solution for the response
density, a fact which would by no means be implied in an
arbitrary choice of potential model. Although we do not make
explicit use of the library of orbits of the final solution in the
present paper, and also no guarantee for the
stability of the models is provided in Kaufmann and Contopoulos
(1996), the self-consistency property suggests that the spiral arms
found in these galaxy models can be stellar dynamically supported.
This conclusion is independent of the theory of the invariant manifolds,
thus the latter theory can be tested against this conclusion.

The relative importance of the various non-axisymmetric components
of the force with respect to the axisymmetric force vary with the
distance from the center, as can be inferred from Figure 2. The bar
contributes to the forcing by both an axisymmetric and a
non-axisymmetric component. The axisymmetric component is found as
the azimuthally averaged radial bar force
\begin{equation}\label{fbarav}
\bar{F}_{bar,r}(r)={1\over
2\pi}\int_0^{2\pi}F_{bar,r}(r,\theta)d\theta~~,
\end{equation}
where $F_{bar,r}=(F_{bar,x}x+F_{bar,y}y)/r$. The total axisymmetric
force is $F_{ax}(r)=F_{disk}(r)+F_{halo}(r)+\bar{F}_{bar,r}(r)$. The
non-axisymmetric bar force at a position $r,\theta$ is the
difference
$\mathbf{F}_{non-ax}=\mathbf{F}_{bar}-\bar{F}_{bar,r}\mathbf{\hat{e}}_r$
where $\mathbf{\hat{e}}_r$ denotes the unit vector in the radial
direction. On the other hand the spiral force corresponding to the
potential can be all considered as non-axisymmetric since the spiral
potential only has a $\cos 2\Phi$ dependent term. Figure 2 shows the
absolute ratios $F_{bar,non-ax}/F_{ax}$ and $F_{spiral}/F_{ax}$ as a
function of the radial distance $r$, along two directions, i.e.,
along the bar's major (dashed  curves) and minor (solid curves)
axes, for all three models.

In model A (Fig.2a) the bar yields the dominant
non-axisymmetric perturbation at all distances up to a zone around
corotation (shown as a vertical dashed line at
$r=6.11\mbox{Kpc}=r_{L1}$). The maximum amplitude of the
non-axisymmetric bar force is 0.32, corresponding to a peak of the
$F_{bar,y}$ curve at $R\simeq 4$Kpc (in all the panels of Fig.2 the
innermost local maxima or minima of the curves $F_{bar,x},F_{bar,y}$
at $R\leq 1$Kpc are artificial, due to the weakening of the
axisymmetric forces which, for a Plummer sphere, are exactly equal
to zero at $R=0$). The inner width of the zone is found by the point
where $F_{bar,non-ax}=F_{spiral}$, which is at a distance $r\simeq
5.75$Kpc. Beyond that distance, the spiral term dominates over the
bar term, reaching a maximum amplitude equal to $0.21$ with respect
to the axisymmetric background. The oscillations of the spiral force
beyond corotation are due to the logarithmic dependence of the
argument $\Phi$ in (\ref{spirpot}) on $r$, a fact causing successive
maxima and minima of the spiral force at successive periods of
length $2\pi$ of the argument $\Phi$. The first maximum, around
corotation, is the most important. The width of the oscillation from
this maximum to the next defines an approximate value of the radial
wavelength of the spiral density wave, which is $\Delta r\simeq
1$Kpc.

In model B (Fig.2b) the maximum amplitude of the
non-axisymmetric bar force reaches the value $0.75$ (for $F_{bar,y}$
at $R\simeq 2$Kpc), implying that the bar is quite strong inside
corotation. Nevertheless, even in this galaxy the spiral force
becomes dominant over the bar's non-axisymmetric perturbation around
and beyond corotation. The zone around the first maximum of the
spiral force defines a radial wavelength of the spiral density wave
$\Delta r\simeq 1$Kpc. The first peak of the spiral force is again
found to be at a distance very close to the corotation radius
$r=r_{L1}=3.43$ and the amplitude of this peak is $0.35$.

Finally, in  model C (Fig.2c) the spiral perturbation near and
beyond corotation reaches such a high amplitude (maximum = 0.57),
that it becomes even stronger than the maximum amplitude of the
bar's perturbation ($\simeq 0.3$ for $F_{bar,y}$ at $R\simeq
3.5$Kpc) which takes place well inside corotation. Furthermore, the
spiral arms are tightly wound (the radial wavelength is estimated as
$\Delta r\simeq0.5$Kpc), and the spiral arms extend to cover about
one azimuthal period $2\pi$. Thus, model C is exceptional and
will not be discussed in detail in the sequel. Only a feature of
this model interesting for dynamics is discussed in Sect.2.3.

\begin{figure}
\centering
\includegraphics[scale=0.7]{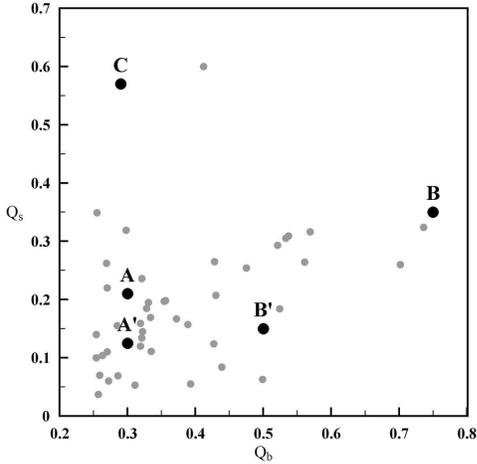}
\caption{The $Q_s$ vs $Q_b$ values in observed SB galaxies (gray
points from the Buta et al. (2005) sample) and our models A, A$'$,
B, B$'$, C (black dots).} \label{}
\end{figure}
The bar-spiral amplitudes of models A and B define strongly
nonlinear models, which are above the average but inside the range
of bar-spiral strengths found by recent observations (Laurikainen \&
Salo 2002; Buta et al. 2005). In the latter works the maximum value
of the ratio of the tangential force versus the radial force for the
bar and spiral components is denoted by $Q_b$ and $Q_s$
respectively. The average values for SB galaxies in the sample of
Buta et al. (2005) are $<Q_b>=0.29$ and $<Q_s>=0.17$. Model A yields
$Q_b=0.3$ and $Q_s=0.21$. Model B yields $Q_b=0.75$ and $Q_s=0.35$,
both values being a factor 1.3 larger than the specific estimates
reported by Buta et al. (2005) for the galaxy NGC1073 ($Q_b=0.56$,
$Q_s=0.26$), to which model B is associated. In order to have a more
representative sample of models in which the theory of the invariant
manifolds is to be tested, two `weak' models, A$'$ and B$'$, are
also considered, which were created by varying some parameters of
models A and B. In model A$'$ the spiral amplitude is $A=1000$, i.e.
half the value of model A (see table I). In model B$'$ the bar mass
is $M_b=0.09$ and the spiral amplitude $A=2500$. Since these changes
are rather arbitrary, there is no guarantee of self-consistency of
the new models. However, a rough criterion of self-consistency
(subsection 2.2) can be established if the pitch angle is also
slightly varied in both models ($i_0=-9^0$ in model A$'$ and
$i_0=-8^0$ in model B$'$). The resulting $Q_b$ and $Q_s$ values are
$Q_b=0.3$, $Q_s=0.125$ for model A$'$ and $Q_b=0.5$, $Q_s=0.15$ for
model B$'$. The spiral strengths of models A$'$,B$'$ are well below
the average of the Buta et al. sample for SB galaxies. Thus, the
mean values $<Q_b>$ and $<Q_s>$ of the four models A,B,A$'$,B$'$
become both representative of the average values found in the
observations. On the other hand, the value $Q_s=0.57$ of model C is
untypical although still in the range of the observations (Figure
3).

\subsection{Phase portraits and invariant manifolds}

The first result of the analysis of the invariant manifolds
can now be demonstrated with the help of Figs.4 to 6. Figure 4a
shows the phase portrait (surface of section $(\theta,p_\theta)$
corresponding to the apocentric positions $\dot{r}=0$,
$\dot{p}_r<0$) in the case of model A, for a value of the Jacobi
constant $E_J=-1.91\times 10^5$, which is close to the value
$E_{J,L1} = -1.915\times 10^5$. Figure 4c shows the same portrait
for $E_J = -1.91\times 10^5$ in a so called `aligned spiral' version
of model A in which the angle $\Phi$ in Eq.(\ref{phasen}) is
replaced by $\theta$ throughout the whole radial extent of the
spiral arms. This means to artificially `align' the spiral arms as
extensions of the bar along the latter's major axis. By this
way we measure the effect of only increasing the amplitude of the
non-axisymmetric perturbation on the form of the invariant
manifolds, while in the original model the manifolds are affected
both by the strength of the non-axisymmetric perturbation and by the
azimuthal displacement of the unstable Lagrangian points with
respect to the bar's major axis.

In Figs.4a,c the points marked PL1, PL2 correspond to the fixed
points of the PL1 or PL2 short-period orbits which are close to the
positions of the unstable equilibria $L_1$, $L_2$. Furthermore, the
thick dots show the intersection of the unstable manifolds ${\cal
W}^U_{PL1}$ and ${\cal W}^U_{PL2}$ with the surface of section. In
order to facilitate the reading of these diagrams, we note that, for
$p_r=0$ (apsides), beyond some radius $r_0\simeq 1$Kpc,
Eq.(\ref{hamgen}) yields that $r$ increases nearly monotonically
with $p_\theta$ in all azimuthal directions of a fixed angle
$\theta$ (a small reversal of this monotonic relation, due to the
non-axisymmetric potential terms, is only observed at angles
$\theta\simeq\pi/2$ and in a small interval of radii, of width
$\Delta r< 0.1$Kpc around $r=5$Kpc; the monotonic relation is
re-established after crossing this interval). Thus, in Figs.4a,c the
semi-plane of the phase portrait with $p_\theta>p_{\theta,L1}$ means
apocentric positions outside corotation, while
$p_\theta<p_{\theta,L1}$ means apocentric positions inside
corotation. Note that in this and in all subsequent plots of phase
portraits the values of $p_\theta$ are normalized with respect to
the value $a^2\Omega_p$, corresponding to the angular momentum in
the rest frame of a circular orbit at a radius $r=a$.

\begin{figure*}
\centering
\includegraphics[scale=0.8]{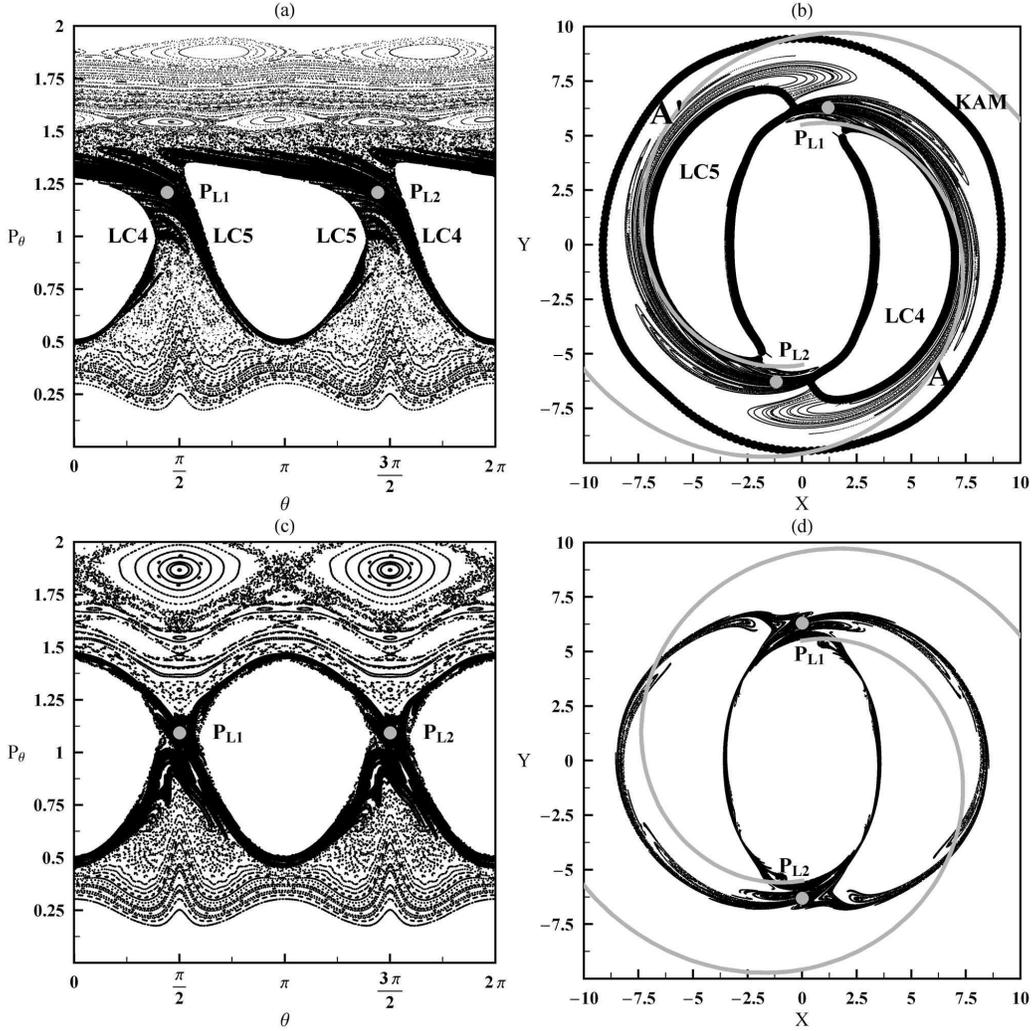}
\caption{(a) Phase portrait near corotation ($E_J=-1.911\times
10^5$) in the case of model A. The thick gray points mark the
position of the fixed points of the PL1 and PL2 orbits. The thick
dark lines are the invariant manifolds ${\cal W}^U_{PL1}$ and ${\cal
W}^U_{PL2}$. (b) Projection of the manifolds of (a) in the
configuration space. The gray spiral curves correspond to the
density maxima of the imposed spiral arms of the full model. The
curves marked LC4,LC5 and KAM are commented in the text. (c), (d)
Same as (a), (b) in the `aligned spiral' ($\Phi=\theta$) version of
model A, in which the minima of the spiral potential term are
aligned with the bar, and $E_J=-1.91\times 10^5$.} \label{}
\end{figure*}
\begin{figure*}
\centering
\includegraphics[scale=0.8]{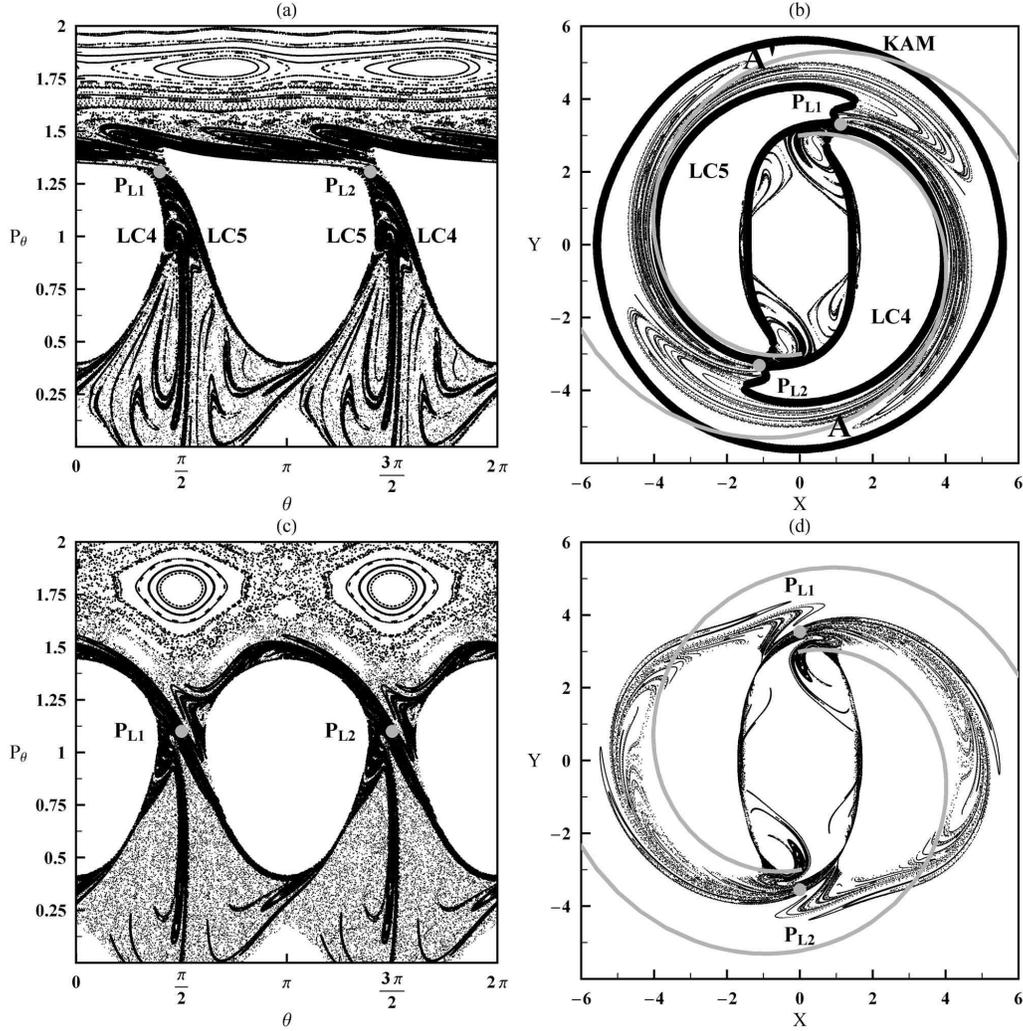}
\caption{Same as in Fig.4, but for model B. The Jacobi constant
is $E_J=-2.98\times 10^4$ in (a) and (b), and $E_J=-2.97\times 10^4$
in (c), (d). } \label{}
\end{figure*}
\begin{figure*}
\centering
\includegraphics[scale=0.8]{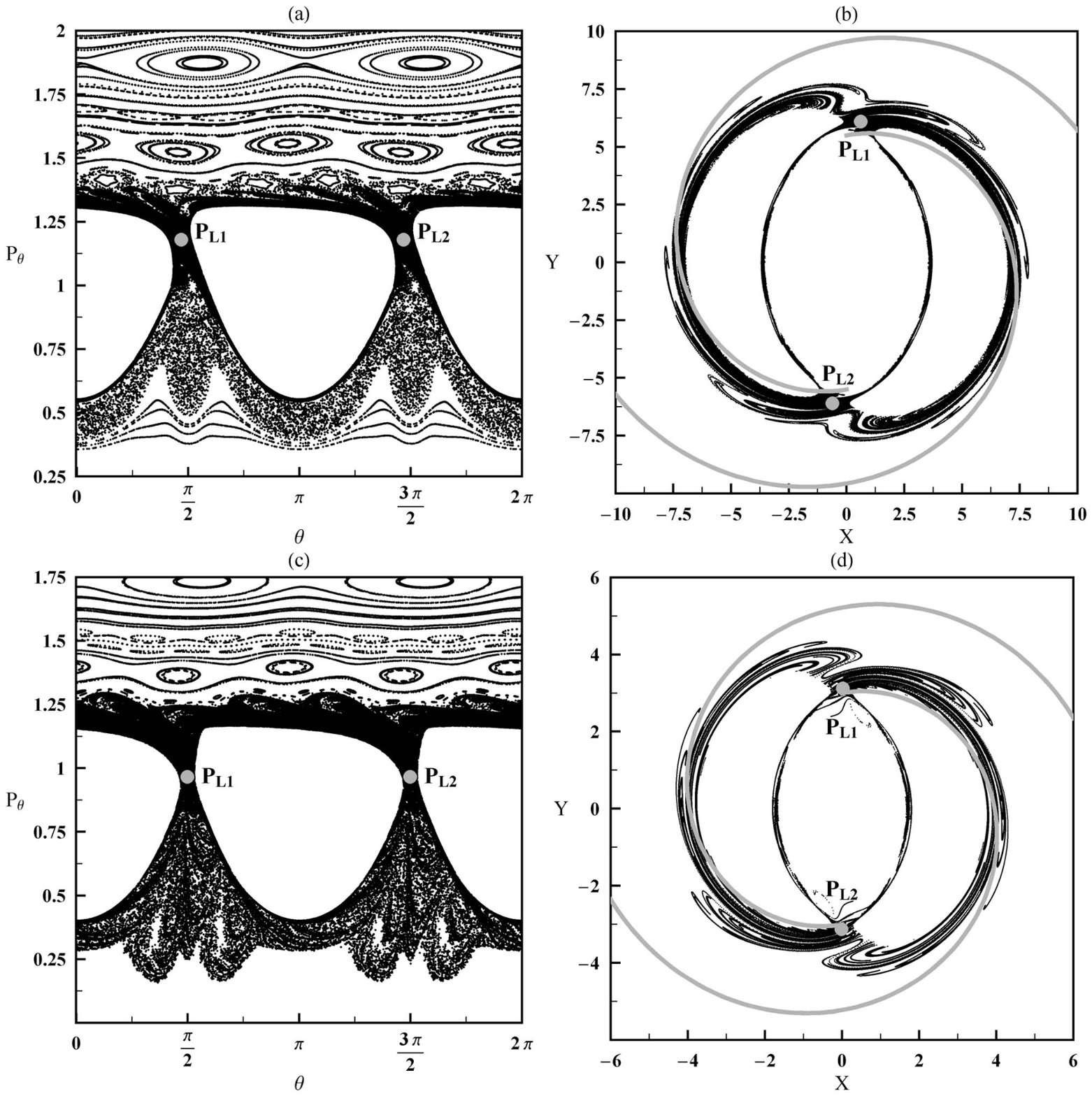}
\caption{(a,b) Same as in Fig.4a,b but for the model A$'$, and
$E_J=-1.91\times 10^5$. (c,d) Same as in Fig.5a,b but for the model
B$'$ and $E_J=-2.77\times 10^4$. } \label{}
\end{figure*}

The main remarks about the comparison of the two phase
portraits are now the following:

- In both portraits chaos is pronounced inside corotation (for
$p_\theta<p_{\theta,L1}$), and the domain of inner invariant KAM
curves is deeply inside the bar (at values of $p_\theta$ about
or below $0.25$). Such extended chaotic domains are responsible for
the termination of the bar.

- Outside corotation (for $p_\theta>p_{\theta,L1}$), a layer of
outer KAM curves has been destroyed in both portraits. This is
caused mainly by the growth of the chaotic layer around the unstable
-6/1 periodic orbit, which produces a resonance overlap with the
chaotic layer of the PL1,2 unstable periodic orbit (negative signs
indicate a resonance outside corotation, for which the motion is
retrograde in the azimuthal direction). As a result, the chaotic
domain extends up to values of $p_\theta\simeq 1.5$, and the first
rotational KAM curves appear a little inside the $-$4/1 resonance.
The domain around the outer Lindblad resonance is almost
entirely filled either by rotational KAM curves or by `resonant'
curves around the $-$2:1 stable periodic orbits. The islands of
stability of the $-$2:1 resonance have a larger width in the
`aligned spiral' model (Fig.4c) because by aligning the spiral
perturbation the amplitude of the total non-axisymmetric
perturbation increases effectively at large distances from
corotation (the width of resonances scales as a power-law of the
non-axisymmetric perturbation (see e.g. Contopoulos 2002).

- The white circular domains devoid of points, embedded in the
chaotic sea of both portraits, correspond to prohibited domains of
motion, for $\dot{r}=0$ and for the selected values of the Jacobi
constant. Such domains exist when $E_J<E_{J,L4}$ (equal to
$E_{J,L4}=-1.873\times 10^5$).

- The most important difference between the two portraits is
that in the case of the true spiral term turned on (Fig.4a) the
prohibited domains lose their azimuthal symmetry with respect to the
values $\theta=0$ (equal to $2\pi(mod 2\pi)$), or $\theta=\pi$
referring to the positions of the stable Lagrangian points $L_4$ and
$L_5$. Such a symmetry is perfect in the aligned spiral case
(Fig.4c). The limiting boundaries of the prohibited domains are
denoted by `LC4', `LC5' in Fig.4. The azimuthal deformation of the
prohibited domains corresponds to an azimuthal deformation of the
associated banana-like prohibited domains appearing in the
configuration space (i.e. the disk plane). These domains are similar
but should not be confused with the domains delimited by the zero
velocity curves of the effective potential in the rotating frame,
i.e. $V_{eff}=V(r,\theta)-\Omega_p^2r^2/2$.

It turns out that the azimuthal deformation of the limiting
boundaries LC4 and LC5 is a crucial difference related to the
morphology of the spiral arms. In both Figs.4a,c the manifolds are
well developed inside and outside corotation (below and above PL1).
In particular, the invariant manifolds form conspicuous lobes and
foldings which are typical of systems having a large degree of the
so-called {\it homoclinic chaos.} However, when the manifolds
are plotted in the configuration space (Figs.4b,d), the azimuthal
deformation of the limiting boundaries LC4 and LC5 in the truly
spiral model (Fig.4b) causes the manifolds to be also azimuthally
deformed in a way so as to closely support the imposed spiral arms
up to an azimuth $\theta'\approx 3\pi/4$ measured clockwise from
either $L_1$ or $L_2$s. The end of this agreement is in two regions
(marked A and A$'$) in which the manifolds ${\cal W}^U_{PL1}$ and
${\cal W}^U_{PL2}$ lose contact from the locus of maxima predicted
by Eq.(\ref{phase}). Beyond this distance, the manifolds exhibit a
typical behavior called a `bridge' in our previous works (Voglis et
al. 2006, Tsoutsis et al. 2008). Namely, the manifold ${\cal
W}^U_{PL1}$ forms an inner spur consisting of a number of lobes
connecting segments of the manifold from A to their continuation,
which starts in the neighborhood of the {\it other} periodic orbit,
PL2. From that point on the oscillations of the manifold ${\cal
W}^U_{PL1}$ start supporting both the spiral arm emanating from
$L_2$ and the chaotic layer marking the border of the bar. When
the manifold is calculated for even longer lengths, we find that the
higher order lobes of the manifold from $L_1$, after reaching the
neighborhood of $L_2$, come again close to $L_1$ in directions
nearly parallel to those formed by the low order lobes of the same
manifold. This causes an enhancement of the density close to $L_1$
along the direction of the imposed spiral arms, i.e., the manifolds
support the self-consistency of the spiral arms.

On the other hand, the manifolds of the `aligned spiral' version of
model A (Fig.4d), calculated up to a length comparable to that of
the manifolds of Fig.4b, show no support of a spiral structure, but
only yield a thick ring-like structure. The thickness of the
manifolds of Figs.4b,d is determined by the degree of chaos in
Figs.4a,c. The degree of chaos is determined by the amplitude of
the non-axisymmetric perturbation. This is expected from dynamical
systems theory, since the overlapping of resonances, which is the
main source of production of chaos, depends on the width of the
different resonant layers near corotation, which, in turn, depends
on only the amplitude of the perturbation.

 A more elaborate analysis (Sect. 2.3) shows that, while the
outermost radial limit of the invariant manifolds is posed by the
existence of absolute barriers, i.e. rotational KAM tori
(marked `KAM' in Fig.4b), more stringent limits are practically
posed by partial barriers, i.e. {\it cantori}, which limit the
diffusion within a chaotic zone. Provided these limits, the
azimuthal deformation of the invariant manifolds is the crucial
factor for the production by them of response spiral arms. This, in
turn, is determined by the form of the limiting boundaries LC4 and
LC5. The theoretical derivation of these boundaries is given in
(Sect.2.3).

Figure 5 shows the same phenomena in the case of model B. The
qualitative resemblance between Figs.4a,b and 5a,b is obvious,
although the azimuthal deformation of the limiting boundaries
LC4 and LC5 is more pronounced in Fig.5b than in Fig.4b. Also
in this model the manifold exhibits a bridge starting at an angle
$\theta'\simeq 3\pi/4$ clockwise from $L_1$ or $L_2$ (points $A$,
$A'$), as well as inner spurs connecting segments of it along both
spiral arms and along the border of the bar. Another feature of
Fig.5b is that the inner branch of the invariant manifold (inside
the bar) is developed in a domain occupying about one fourth of the
total extent of the bar. This implies that a substantial part of the
bar in the domain near corotation is supported by chaotic orbits. In
fact the non-axisymmetric forcing in model B is much stronger inside
corotation than in model A, a fact causing the destruction of
all the inner KAM curves down to $p_\theta=0$ (Fig.5a). Such a
type of chaos may lead to a number of observational consequences,
photometric and kinematic, a list of which have been enumerated by
Grosb{\o}l (2003). Finally, the azimuthal deformation of the maxima
of the spiral term with respect to the bar's major axis also turn
out to be the crucial factor for the production by the manifolds of
response spiral arms. In fact, by comparing Figs.5a,b with the
respective figures in the `aligned spiral' version of model B (
Figs.5c,d) we see that the manifolds in the latter case
present some asymmetry as well as a large thickness, due to the high
value of the non-axisymmetric perturbation, but they still largely
deviate from the spiral pattern (gray locus), which was closely
followed by the manifolds of the non-aligned model (Fig.5b).

Figure 6 shows the phase portrait structure near corotation in
the `weak spiral' models A$'$ (Fig.6a) and B$'$ (Fig.6b). The
invariant manifolds ${\cal W}^U_{PL1}$ and ${\cal W}^U_{PL2}$ are
also plotted, and the counterparts of these plots in the
configuration space are shown in Figs.6b and 6d respectively. As
expected, in both models chaos is considerably reduced with respect
to the strongly nonlinear models A,B, and it is only limited in a
narrow zone in the corotation region. Further away, the phase space
is filled by invariant tori which occupy most of the phase space
volume already at the -4:1 resonance.

The $Q_s$ value of model A$'$ is $Q_s=0.125$, and this is its only
difference with respect to model A, which has $Q_s=0.21$. The
thickness of the invariant manifolds is thus reduced with respect to
the thickness of the manifolds of model A (compare Figs.6b and 4b).
However, the azimuthal deformation of the manifolds is still large
enough to fit the locus of the imposed spiral arms up to an angle
$\theta'=\pi/2$ clockwise from $L_1$ or $L_2$, i.e. the manifolds
support quarter turn spiral arms. In the case of model B$'$ (Fig.6d)
we have $Q_s=0.15$, which is close but still below the average value
$<Q_s>=0.17$ of the Buta et al. (2005) sample. At the value
$Q_s=0.15$ the region of homoclinic chaos formed by the lobes of the
manifolds near $L_1$ or $L_2$ is already well developed, and the
`inner spurs' are clearly distinguishable. In fact, in model B$'$ a
small adjustment of the pattern speed ($\Omega_p=30.5$ instead of
$32.5$ Km/sec/Kpc, as was in model B) yielded the best fit of the
invariant manifolds to the imposed spiral arms. Such a fit can now
be considered as a rough criterion of self-consistency. We see that
the formation of bridges in the manifolds of Fig.6d result in that
the higher order lobes of the manifold make oscillations which
enhance the density along the manifolds' unstable directions all the
way from $L_1$ or $L_2$. Thus the manifolds support, again, the
imposed spiral arms in a self-consistent way. In fact, since the
overall thickness of the domains covered by the invariant manifolds
increases, the manifolds support the spiral structure up to an angle
$\theta'$ larger than $\pi/2$, i.e. along a length larger than in
model A$'$.

In conclusion, there is a clear morphological continuity of the
structures produced by the invariant manifolds, from rings to
quarter turn spirals, and then to fully developed spiral arms, as
the value of $Q_s$ increases. The azimuthal displacement of the
unstable Lagrangian points is responsible for the manifolds
producing a spiral-like response, while the increase of the
amplitude $Q_s$ extends the support of the spiral structure to
higher angles $\theta'$. Such a morphological continuity of the
invariant manifolds is suggestive of it being a real morphological
feature of barred galaxies.

\subsection{Theoretical modeling}

\begin{figure*}
\centering
\includegraphics[scale=1]{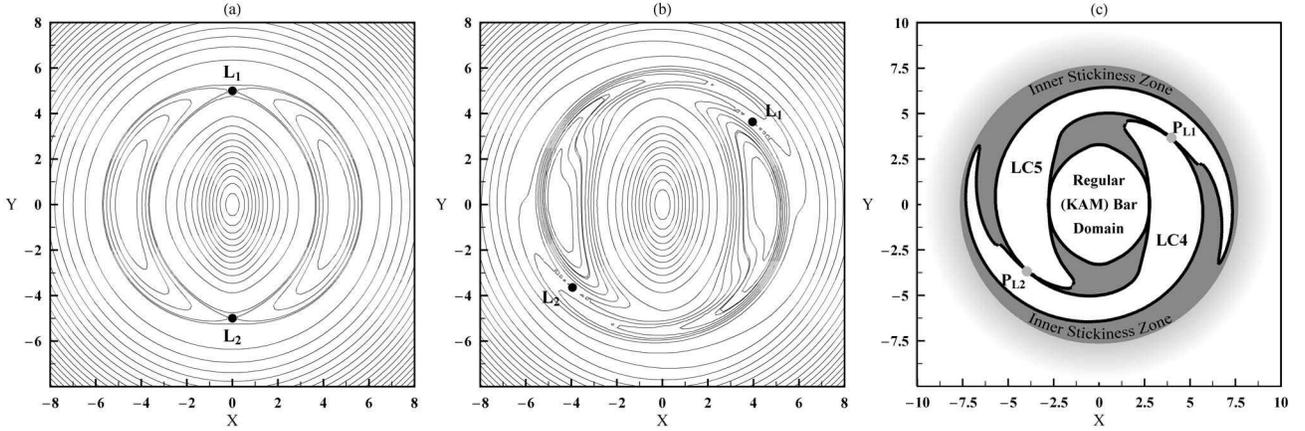}
\caption{The curves of zero velocity (equipotential curves of
the effective potential $V_{eff}$) shown schematically when the
non-axisymmetric perturbation is (a) aligned to the bar, and (b) has
a spiral component. (c) A schematic figure of the basic theory
according to which the structure of the chaotic phase space in the
corotation region supports of trailing spiral arms. The spiral arms
are confined in an inner stickiness zone (dark) delimited by cantori
which are the remnants of rotational KAM tori outside corotation.
Inside the bar, the same chaotic zone is delimited by rotational KAM
tori (regular bar domain). The chaotic zone between the curves LC4
and LC5 and the boundary of the stickiness zone has a spiral shape,
due to the deformation of the limiting curves LC4 and LC5 caused by
the azimuthal displacement of L1 and L2 with respect to the bar's
major axis. } \label{}
\end{figure*}
A first estimate of the azimuthal deformation of the limiting
boundaries LC4 and LC5 can be done by calculating the azimuthal
displacement of the unstable Lagrangian points $L_{1,2}$ when
$V_{spiral}$ is turned on (Figs.7a,b, schematic). This can be judged
from the form of the equipotential curves (called hereafter the
`curves of zero velocity', CZV) in the rotating frame, i.e., the
level curves of
\begin{equation}\label{vrot}
V_{eff}=V(r,\theta)-\Omega_p r^2/2 = E_J~~.
\end{equation}
For a logarithmic spiral, the azimuthal shift of $L_1$ is given
by
\begin{equation}\label{deltatheta}
\Delta\theta ={1\over\tan i_0}\bigg|\ln{r_{L_1}\over a}\bigg|~~
\end{equation}
where $\Delta\theta$ is taken clockwise from the bar's major axis.
The azimuthal deformation of the CZVs takes place in a narrow zone
around corotation (Fig.7b), while, for $r$ small, the curves
remain practically aligned to the bar. This is because the spiral
potential term ($\ref{spirpot}$) (with $\Phi$ given by
Eq.(\ref{phasen})) adds practically a $\cos 2\theta$
contribution to the bar's term at radial distances $r<a$. The stable
Lagrangian points $L_{4,5}$ are shifted counterclockwise with
respect to the bar's minor axis, although by a smaller angle than
that of the unstable points $L_1$, $L_2$.

The equipotential curves of $V_{eff}$  do not provide the strictest
limit of allowed motions on the {\it apocentric surface of section}
$(\theta,p_\theta)$, or $(\theta,r)$, for $p_r=0$, $\dot{p}_r<0$.
The form of the limiting curves in the configuration space is
shown schematically in Fig.7c (curves LC4 and LC5, encircling the
stable Lagrangian points $L_4$, $L_5$). These curves are derived by
the request that, for any fixed angle $\theta$ and angular momentum
$p_\theta$, the curve of the function
\begin{equation}\label{efpot}
\bar{V}(r;\theta,p_{\theta})={p_\theta^2\over 2r^2}-\Omega_p
p_\theta + V(r,\theta)
\end{equation}
be tangent to the line $\bar{V}=E_J$ at the point where
$\bar{V}=E_J$  has a global minimum. The minimum is calculated by
the root $r=r_{m}(\theta,p_\theta)$ of
\begin{equation}\label{discr}
{\partial \bar{V}\over\partial r}=-{p_\theta^2\over r^3}+ {\partial
V\over\partial r}=0~~.
\end{equation}
The limiting values of $p_\theta$, for fixed $\theta$, are then
found via the roots for $p_\theta$ of
$$
\bar{V}(r_{m}(\theta,p_\theta),\theta,p_\theta)=E_J~~.
$$
The limiting curves LC4 and LC5 in the configuration space, derived
from Eqs.(\ref{efpot}) and (\ref{discr}) are outside the limiting
curves provided by the CZVs defined through Eq.(\ref{vrot}). This is
due to the fact that these limits now refer to $\dot{r}=0$, while in
the case of Eq.(\ref{vrot}) the total velocity in the rotating frame
is equal to zero (details are given in Appendix A). By virtue of
these facts, the curves LC4 and LC5 exhibit also a spiral-like
azimuthal deformation, which is even more pronounced than that of
the CZVs.

The gray domain in Fig.7c shows the permissible apocentric
positions of the orbits under a fixed value of the Jacobi constant
close to the corotation value. The interior gray domain between LC4
and LC5 roughly marks the extent of the bar. On the other hand, the
positions of the PL1 and PL2 points are in very narrow strips of
permissible apocentric positions separating the inner right part of
the LC4 curve from the outer right part of the LC5 curve and vice
versa. The invariant manifolds $W^U_{PL1,2}$ emanating from these
points necessarily follow the narrow strips leading to the outer
gray domain, thus they yield locally the form of spiral arms.

\begin{figure*}
\centering
\includegraphics[scale=0.8]{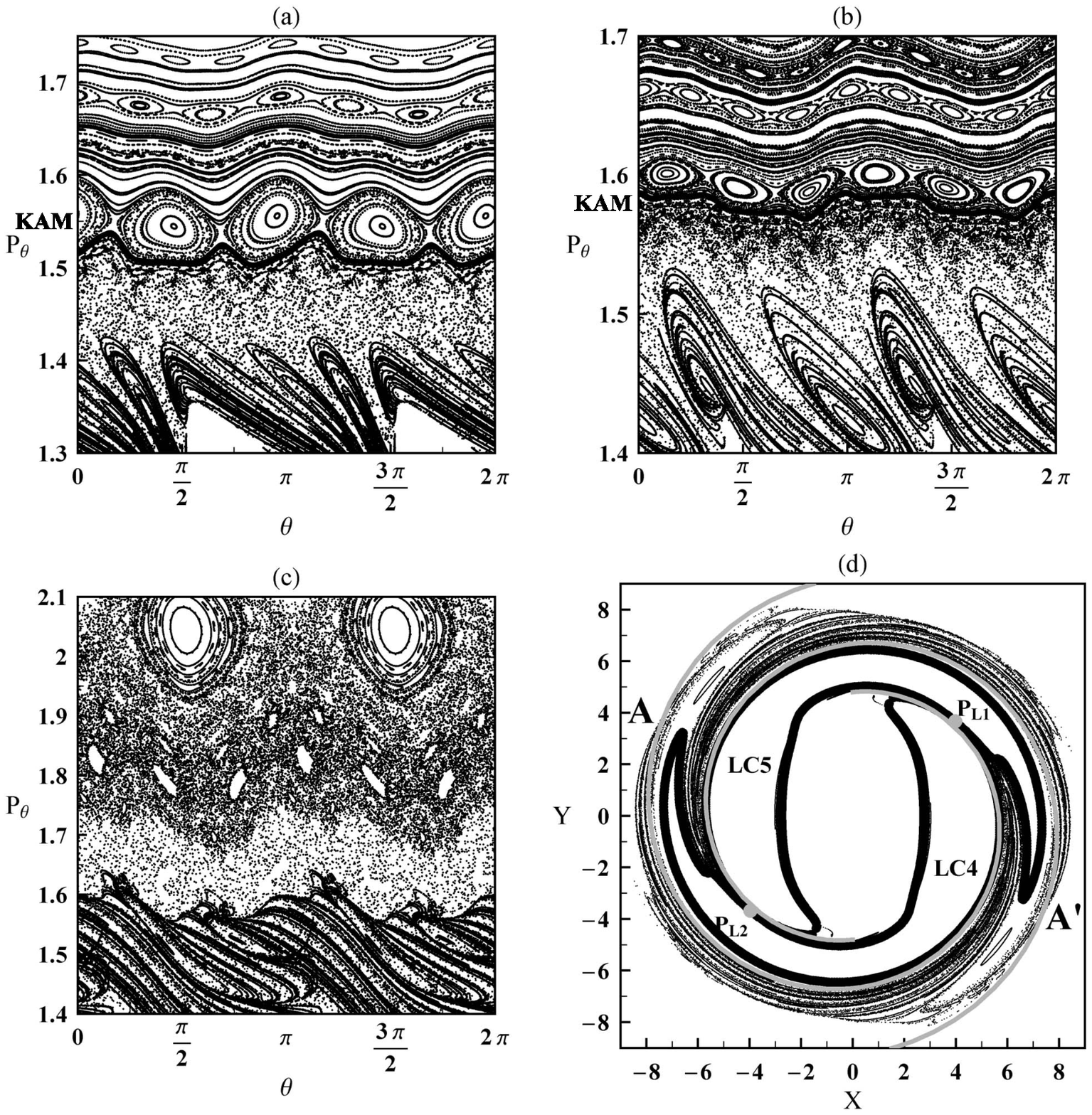}
\caption{ The resonant phase space structure in the corotation
region in the cases of (a) model A with $E_J=-1.911\times 10^5$, (b)
model B with $E_J=-2.98\times 10^4$, and (c) model C with
$E_J=-2.912\times 10^5$. The invariant manifolds of the PL1 and PL2
orbits are over-plotted (thick lines). The bold rotational KAM
curves near $p_\theta=1.5$ in (a) and $p_\theta=1.57$ in (b) are
very close to the inner boundary of the regular domain beyond
corotation. In (c) there is no such boundary at least up to the
outer Lindblad resonance. The projection of the rotational curves of
(a) and (b) marked KAM, as well as of the curves LC4 and LC5, in the
configuration space, yields the respective outer closed curves in
Figs.4b, 5b. (d) Same as in Fig.4b but for the model C, and Jacobi
constant $E_J=-2.912\times 10^5$} \label{}
\end{figure*}

The evolution of the invariant manifolds further away from PL1 or
PL2 is determined by the resonant structure in the outer corotation
zone. The existence of many resonances accumulating in a narrow
range of distances near the corotation radius causes a chaotic layer
in this region, formed by the mechanism of resonance overlap.
Figure 8 makes a zoom to the phase portraits of the models
considered, in order to demonstrate the relevant phenomena.
Figs.8a,b are zooms to the phase portraits of Figs.4a, and 5a,
referring to the models A and B. In both cases we find a chaotic
layer extending up to $p_\theta \simeq 1.5$, which is delimited by a
rotational KAM torus at $p_\theta=1.5$ (marked KAM). This torus is
just below the -4/1 resonance, which is stable at the value of the
Jacobi constant $E_J=-1.911\times 10^5$. On the other hand, most
islands of stability of resonances $-m/1$, with $m>4$, which are
closer to corotation,  have been destroyed. Their destruction is
followed by the destruction of KAM tori with irrational rotation
numbers, which, according to the standard theory, are transformed
into cantori. Such cantori limit the chaotic flux through their
gaps, and this fact causes some stickiness in a zone very close to
the PL1 and PL2 fixed points. Stickiness phenomena of this type have
been explicitly demonstrated and studied in simple models of the
dynamical systems theory (see e.g. Efthymiopoulos et al. 1997;
Contopoulos et al. 1999; Contopoulos \& Harsoula 2008), and they
have also been observed in our $N$-Body simulations of barred-spiral
galaxies (Tsoutsis et al. 2008). The main outcome of these studies
is that cantori in a large chaotic sea act as partial barriers
slowing down considerably the escape (or diffusion) of the chaotic
orbits with initial conditions along or near an invariant manifold.

In the case of the manifolds ${\cal W}^U_{PL1,2}$ plotted in
Figs.8a,b, which are calculated from 21 iterations of an initial
segment of length $ds=10^{-4}$ close to PL1 or PL2, we see that the
manifolds fill only partially the chaotic domain up to the torus
marked KAM. The inner dark region covered by the first iterations of
the invariant manifolds defines a domain called `inner stickiness
zone', the projection of which in the configuration space is shown
schematically as a dark gray domain in Fig.7c. The remaining part of
the chaotic domain up to the curve marked `KAM' corresponds
essentially to the light gray domain of Fig.7c. This domain is
eventually covered by the invariant manifolds after a very large
number of iterations. For example, the manifolds of Figs.8a,b have
not yet reached the curve KAM after about 50 iterations, which in
both models correspond to about 60 pattern rotation periods. Thus,
during all this time interval the manifolds support a spiral
structure.

The stickiness phenomena keep playing a significant role even when
the spiral perturbation is pushed to untypically high values. For
example, in model C, (Figs.8c,d) the spiral strength is
$Q_s=0.57$, and under such a high value all the rotational KAM
curves are destroyed, at least up to the outer Lindblad resonance.
Then, while in principle there is no absolute barrier to chaotic
diffusion up to very large distances from corotation, a plot of the
invariant manifolds (Fig.8c) shows that these manifolds
exhibit again stickiness phenomena, and they practically remain
confined for very large times below the $-4/1$ resonance (which is
still stable at the value of the Jacobi constant $E_J=-2.912\times
10^5$, yielding four tiny islands embedded in the large chaotic sea
of Fig.8c near the level $p_\theta=1.6$. This results in that
the manifolds in the configuration space (Fig.8d) yield the
form of tightly wound spiral arms. In this particular example, the
domain covered by the invariant manifolds practically coincides
with the `inner stickiness domain' of Fig.7c.

\section{Spiral arms as the response of invariant manifolds to bars}
\begin{figure*}
\centering
\includegraphics[scale=1]{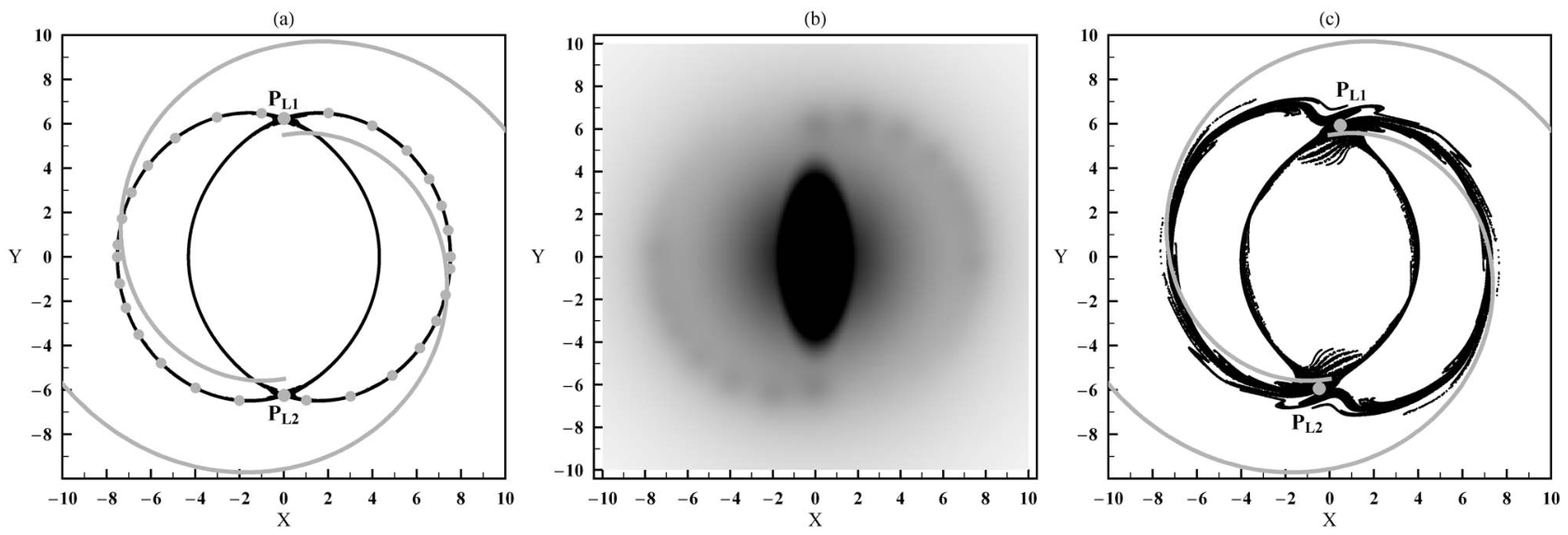}
\caption{A `spiral response' model based an initially `pure
bar' version of model A. (a) Manifolds of the pure bar case for
$E_J=-1.90\times 10^5$ (very close to the $E_{J,L1}$ value. These
manifolds are populated by `point masses' represented as Plummer
spheres centered at the positions indicated by the the gray thick
dots (see text for details). (b) Gray scale mapping of the
surface density produced by the mass distribution of (a). (c)
Response manifolds calculated in the new potential corresponding to
the mass distribution of (a). } \label{}
\end{figure*}
\begin{figure*}
\centering
\includegraphics[scale=0.8]{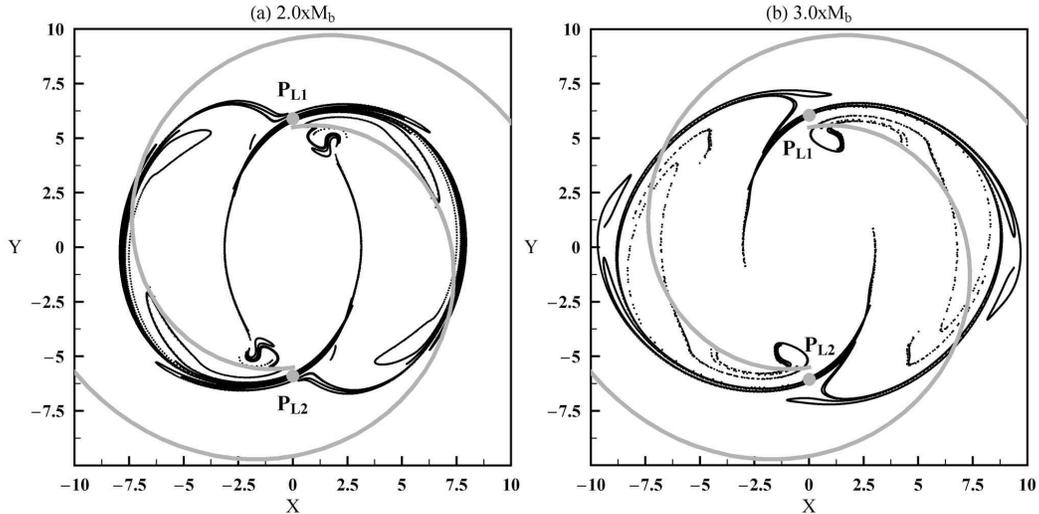}
\caption{Unstable invariant manifolds of the PL1 and PL2 orbits in a
pure bar version of model A in which the bar's mass and pattern
speed are altered with respect to the reference values
$M_b=1.5\times 10^{10} M_\odot$, $\Omega_p=43.6$Km/sec/Kpc so as to
keep corotation at a fixed distance. (a) $M_b'=2M_b$, $\Omega_p'=
1.26 \Omega_p$, $E_J=-2.19\times 10^5$. (b) $M_b'=3M_b$, $\Omega_p'=
1.38 \Omega_p$, $E_J=-2.40\times 10^5$. } \label{}
\end{figure*}
\begin{figure*}
\centering
\includegraphics[scale=0.8]{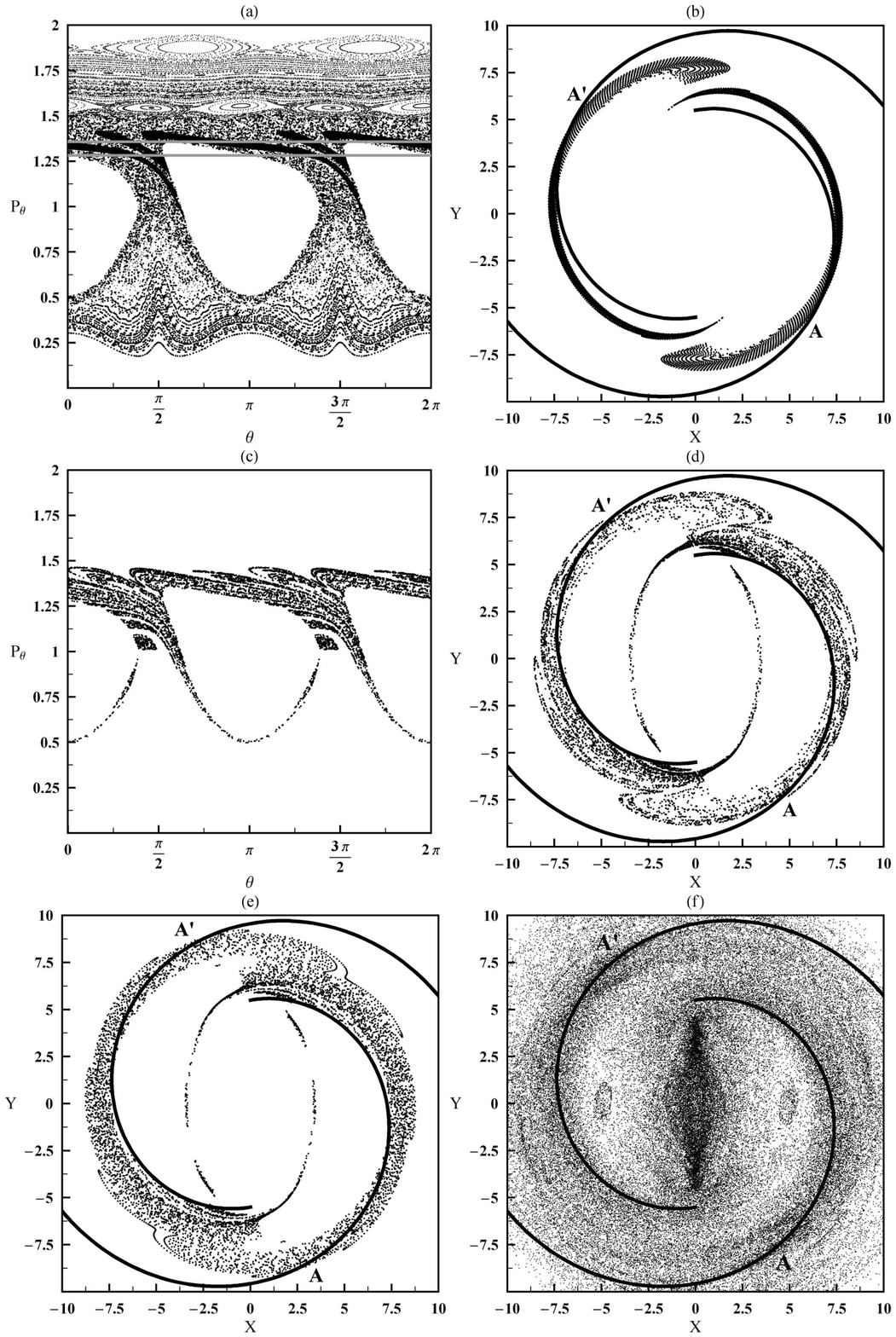}
\caption{Model A for the value of the Jacobi constant
$E_J=-1.911\times 10^5$. (a) The two gray horizontal lines define a
zone of initial conditions around the circular orbit of the
unperturbed (axisymmetric) potential. The thick black dots are the
images of these initial conditions after one iteration of the
Poincar\'{e} map. The background thin points show the underlying
phase space structure. (b) The first image of the Poincar\'{e} map
of the initial conditions specified in (a) as it appears in the
configuration space. After only one iteration this set of points
already acquires the form of the invariant manifolds, i.e., it
yields trailing spiral arms. The projection of the tenth iterates of
the same points on the surface of section and on the configuration
space are shown in (c) and (d) respectively. (e) Same as in (d) when
the initial conditions are taken in a zone four times as wide as the
zone of (a). (f) A full `response model' for the same galaxy
produced according to the methodology suggested in Patsis (2006).
The black curves represent the minima of the imposed spiral
potential.} \label{}
\end{figure*}

One immediate consequence of the analysis of the previous sections
is that one cannot induce the morphology of the spiral arms,
corresponding to a particular morphological type of bar, by
calculating the invariant manifolds of the PL1 and PL2 families in
only a pure bar potential. In fact $V_{spiral}$ is most important
near corotation and it must be taken into account self-consistently
in all studies related to the morphology of the spiral arms via the
calculation of invariant manifolds. This result is in agreement and
probably provides a dynamical basis for understanding the results of
both particle and hydrodynamical simulations (Lindblad et al. 1996;
Aguerri et al. 2001) which have reported the inefficiency of
simulations of pure bars to reproduce a spiral structure.

On the other hand, the theory of the invariant manifolds suggests
that the spiral arms are linked dynamically to the bar. A plausible
scenario for establishing such a link is one in which the bar
initiates the process of a spiral response, which is then enhanced
self-consistently by the growing contribution of the spiral
potential.

The question addressed below is whether, on the sole knowledge of
the gravitational potential and pattern speed of a pure bar model,
such a process can be modeled via the theory of the invariant
manifolds. We construct such an iterative `spiral response' model as
follows:

i) We first calculate the invariant manifolds produced by the pure
bar model.

ii) We assume that the invariant manifolds produced by the pure bar
alone `trigger' the formation of a ring-like or spiral pattern by
attracting matter along the invariant manifold. The density
gradient along the manifold cannot be uniform, since (a) the speed
of chaotic diffusion is smaller close to $L_1$ or $L_2$ than far
from these points, and (b) the higher order lobes of the invariant
manifolds return to the neighborhood of $L_1$ and $L_2$ (Sect.2.2).
In order to model the mass distribution along the invariant
manifolds, we consider a number $N_p$ of small Plummer spheres, of
mass $m_p$ and softening radius $r_p$, placed along the invariant
manifold of the pure bar case (Fig.9a) with a linearly
decreasing mass from $L_1$ or $L_2$ counterclockwise, namely the
mass of the i-th particle is given by:
\begin{equation}\label{mp}
m_p(i)=\frac{2M_s}{N_p}-\frac{2M_s}{\left(\frac{N_p}{2}+1\right)N_p}i
\end{equation}
with $i=1,2,...,N_p/2$ from $L_1$ to $L_2$ counterclockwise and similarly
along the symmetric manifold from $L_2$ to $L_1$. $M_s$ is an estimate
of the total mass on the spiral arms (see below). The choice of a linear
mass decrease as in (\ref{mp}) along the response spirals is rather arbitrary
and it does not follow directly from the theory of the invariant manifolds.
However, it does capture the essential feature that the density of points
should in general decrease along the unstable manifold as we recede from
the unstable periodic orbit.

In the simulation of Fig.9, we start from the invariant manifolds of the
pure bar version of model A, and set $r_p=0.5$, $N_p=30$, and the
mass of each particle fixed so that the total mass of all the
particles is equal to $M_s$ given by
\begin{equation}\label{mspir}
M_s={1\over 2}\int_{r_1}^{r_2}\int_0^{2\pi}\int_{-\infty}^{\infty}
|\rho_s(r,\theta,z)|rdr d\theta dz
\end{equation}
where the quantity $\rho_s(r,\theta,z)$ is an approximate expression
for the density perturbation corresponding to the potential
(\ref{spirpot}) given by the WKB ansatz:
\begin{equation}\label{rhospir}
\rho_s(r,\theta,z)=-{|k\Delta|(|k\Delta|+1)\over 4\pi G\Delta^2}
V_s(r,\theta) \mbox{sech}^{2+|k\Delta|}(z/\Delta)
\end{equation}
where $k=2/(r\tan i)$, according to the formula given by Vandervoort
(see Contopoulos and Grosb{\o}l 1988; the used values of $\Delta$
are given in Table 1). Thus, $M_s$ represents an estimate of the
total mass contained in the spiral arms of model A. A gray
scale plot of the total surface density corresponding to the above
mass distribution is shown in Fig.9b.

iii) The gravitational potential is calculated anew taking into
account the `response' potential produced by the Plummer masses
positioned as described in step (ii). This also yields a new
position of the Lagrangian points $L_1$, $L_2$ as well as a new form
of their invariant manifolds. The procedure should be repeated until
convergence of the positions of the Lagrangian points and of their
manifolds towards a final form is obtained. In practice, we
find that just one iteration suffices to obtain invariant manifolds
which fit the imposed spiral perturbation up to an azimuth
($\theta'\simeq \pi/2$). In particular, the manifolds of Fig.9c
cover a much thicker radial domain than those of the initial `pure
bar' model (Fig.9a), and they also exhibit an azimuthal deformation
following from the non-uniform distribution of mass shown in
Fig.9b.

The present method can in principle be used to explore the sequence
of different morphological types of spiral arms obtained by the
method of the invariant manifolds applied to a family of pure bar
potentials, when the bar parameters, i.e., bar strength and pattern
speed are altered. Such a type of study was undertaken by
Romero-Gomez et al. (2007), without taking, however, into account
the effects of the self-consistent spiral response. For our
adopted models and parameters, we find that when this effect
is ignored, a kind of spiral response can only be
produced by the low-order lobes of the invariant manifolds, but i)
the bar strength has to be pushed to very high values, and
ii) the spiral form disappears when the manifold is calculated for
longer length. An example is given in Fig.10, referring again
to the `pure bar' version of model A. The invariant manifolds are
calculated after altering the bar parameters with respect to the
values given in Kaufmann \& Contopoulos (1996). Figure 10 shows two
different choices of parameters, in which the bar is given a mass
equal to (a) two and (b) three times the mass value in Kaufmann \&
Contopoulos (1996). The pattern speed was altered accordingly so as
to keep corotation at a fixed distance. As the bar amplitude
increases, the first lobes of the invariant manifolds become more
open and they yield a gradual transition from a ring (Fig.10a) to a
spiral pattern (Fig.10b). However, the spiral pattern in
Fig.10b does not fit the self-consistent spiral pattern of model A.
Furthermore, the change of the bar strength required in order to
produce this result constitutes a large deviation from the
parameters of the original model which guaranteed self-consistency.
This can be probably improved by choosing a lower $n-$value of the Ferrers
bar and/or including the effects of a more rectangular bar outline.

A different methodology to produce response models of barred-spiral
galaxies has been proposed by Patsis (2006). As already emphasized
in Tsoutsis et al. (2008), the theory of the invariant manifolds
explains many features of such response models. Patsis (2006)
considers an ensemble of particles with initial conditions on
circular orbits of the axisymmetric part of the potential, placed
uniformly on the disk at radii corresponding to Jacobi constants up
to the $L_4$ value of the full potential. Then, by softly
introducing the non-axisymmetric part of the potential (the
transition time is a few pattern periods), the particles'
distribution changes due to both the adiabatic change of the
potential and phase mixing. As a result, the particles finally
settle to orbits supporting both the bar and the spiral arms.

Inside the bar, the particles of such response models are captured
mostly in stable resonances belonging to a branch or bifurcation of
the $x_1$ family, which is the continuation of the family of
circular orbits of the axisymmetric potential on which all the
initial conditions lie. However, near corotation and beyond chaos is
prominent. In this case the theory of the invariant manifolds
explains the capture of the particles in orbits supporting the
spiral arms. In our specific models, one can see this effect by
taking initial conditions $(\theta,p_\theta)$ on the surfaces of
section such as those of Figs.4a and 5a such that these initial
conditions a) belong to a chaotic domain of their respective
surfaces of section, and b) determine circular orbits when all the
non-axisymmetric part of the potential is turned off (the monopole
contribution of the bar, obtained by averaging radial forces with
respect to all possible azimuths, is taken into account in this
calculation). The locus of all these initial conditions is a
straight line $p_\theta=const$ on the surface of section for a given
Jacobi constant. In practice, we take a narrow zone of some width
around such a line, i.e., allow also for a small value of the
epicyclic action around the circular orbits of the axisymmetric
model.

Figure 11 shows the result of running these initial conditions in
the case of model A, with the non-axisymmetric part of the potential
being turned on from the start, and for one iteration (Figs.11a,b,
corresponding to 1.2 pattern rotation periods), or ten iterations
(Figs.11c,d, corresponding to 12 pattern rotation periods), for the
value of the Jacobi constant $E_J=-1.911\times 10^5$. The initial
conditions correspond to a zone in Fig.11a between the two gray
horizontal lines, that is we take 10000 points uniformly distributed
within the intersection of the zone with the permissible domain of
motion. The first iterates of these points are shown with dark thick
points in the same figure, while the large number of small dots
illustrate the overall structure of the phase portrait at the chosen
value of $E_J$. The corresponding figure in configuration plane is
shown in Fig.11b. An obvious conclusion from Figs.11a,b is that,
already after one iteration, the initially straight zone of initial
conditions in the surface of section is deformed so as to closely
follow a pattern induced essentially by the invariant manifolds
${\cal W}^U_{PL1,2}$ (compare Figs.11a,b with Figs.4a,b). This
phenomenon is repeated at subsequent iterations, so that after ten
iterations (Figs.11c,d) the set of all points describes a pattern
nearly coinciding with that of the manifolds ${\cal W}^U_{PL1,2}$,
but over a larger length of the latter. This picture does not change
qualitatively if the zone of initial conditions is taken to have a
width four times as large as in Fig.11a. The resulting response of
the orbits in this case is shown in Fig.11e, and the association of
this with the dynamics of the invariant manifolds is still quite
clear.

This behavior follows from a well known `mixing property' of chaotic
dynamical systems, namely that any small and compact ensemble of
initial conditions embedded in a large chaotic domain is deformed in
subsequent Poincar\'{e} mappings so as to follow the unstable
invariant manifolds of the main families of unstable periodic orbits
located in the chaotic domain, while preserving its measure in the
same domain (see Contopoulos \& Harsoula (2008) and references
therein for a detailed exploration of this phenomenon in the case of
the `standard map'). In the case of barred galaxies, this type of
response of the chaotic orbits to the invariant manifolds generates
patterns such as those of Figs.11b,d i.e., spiral arms.

It should be stressed, however, that Figs.11b to 11e were calculated
with the full non-axisymmetric potential, i.e.,
$V_{bar}+V_{spiral}$, turned on from the start, i.e., without some
initial transient time of growth of the non-axisymmetric part, and
for just one value of the Jacobi constant. Figure 11f shows the full
`response model' obtained precisely via the method suggested by
Patsis (2006). We thus take the initial conditions of $10^5$
particles on circular orbits of the axisymmetric part of the
potential and on part of a uniformly populated disk, with distances
in a range corresponding to the whole range of Jacobi constant
values up to $E_{J,L4}$ of the full potential. The
non-axisymmetric part of the potential was now introduced softly (in
two pattern periods), and the particles were found to settle to a
nearly invariant distribution in the configuration space after about
15 pattern periods. Figure 11f shows this distribution at a time
corresponding to the 25th pattern period. Clearly, the particles
have settled to orbits supporting both the bar and the imposed
spiral arms. But the most interesting feature of this distribution
is that the maxima of the density of the response spirals depart
from the maxima of the imposed spiral at nearly the same points
(points $A$ and $A'$) where this happens for the invariant manifolds
of Fig.4b, which are also essentially traced by the points in
Figs.11b,d, or e. The interest of this result lies in that the
initial conditions of the response model of Fig.11f are selected
from a nearly homogeneous distribution in space, and thus they are
by no means associated with the invariant manifolds. In this
respect, the invariant manifolds play for chaotic orbits a role
similar to that of stable resonances for regular orbits, i.e. the
manifolds are able to capture the chaotic orbits in their
neighborhood and to create response spiral arms.

\section{Conclusions}

In the present paper we examined the applicability of the theory of
the invariant manifolds emanating from the unstable short period
orbits around the Lagrangian points $L_1$ and $L_2$ of a
barred galaxy in the self-consistent models of Kaufmann \&
Contopoulos 1996) which are rough models representing some features
of real barred-spiral galaxies. We also tested the theory in
weak or mean spiral versions of these models. Our conclusions are
the following:

1) When both the bar and spiral components of the self-consistent
models are taken into account, the projection of the invariant
manifolds on the configuration space produces a pattern that follows
closely the imposed spiral pattern of the self-consistent model.

2) The addition of the spiral potential produces two effects: a) it
shifts the position of the unstable Lagrangian points $L_1$, or
$L_2$, both radially and angularly, and b) it enhances chaos
locally, in a zone around corotation, due to the increase of the
amplitude of the non-axisymmetric perturbation. The azimuthal
displacement of the Lagrangian points is the most important
factor for the manifolds to obtain a spiral form. In artificial
models in which the whole non-axisymmetric perturbation is `aligned'
to the bar (e.g. by setting the phase of the spiral term equal to
$\Phi=\theta$), the manifolds yield ring rather than spiral
structures.  If the amplitude of the perturbation is pushed to
very high values, the lowest order lobes of the invariant
manifolds can determine spiral patterns. However, we find that
such patterns i) do not fit the spiral patterns of the
self-consistent models, and ii) disappear after a few iterations of
the calculation of the manifolds. This could probably be improved
also by considering a lower $n-$value of the Ferrers bar, yielding a more
gradual decrease of the bar force beyond the bar's end, and/or
rendering the bar's outline more rectangular.

3) We construct a simple theoretical model yielding the
boundaries of the apocenters of the chaotic orbits in the
configuration space. Since the manifolds are developed within these
boundaries, we demonstrate that they necessarily take the form of
spiral arms, provided that the azimuths of the Lagrangian points
$L_1$, $L_2$ are shifted with respect to the bar's major axis.  The
outermost radial limits of the manifolds are posed by rotational KAM
curves outside corotation. However, in practice there are more
stringent limits posed by the existence of cantori limiting the
chaotic diffusion. The cantori define a narrow stickiness zone
beyond corotation. The stickiness causes confinement of the
invariant manifolds inside this zone for times comparable to the
age of the galaxy.

4) A simple iterative method is proposed to calculate `spiral
response' models on the basis of the theory of the invariant
manifolds, starting from the invariant manifolds of a pure bar
model. We also discuss the relevance of the response models of
Patsis (2006) to the theory of the invariant manifolds. Following a
general property of the Hamiltonian flow in a connected chaotic
domain, a small sub-domain of initial conditions embedded within the
chaotic domain yields successive Poincar\'{e} maps following closely
the form of the invariant manifolds of the main families of unstable
periodic orbits in the same domain. The consequences of this
effect are tested by calculating a full response model via the
method proposed by Patsis (2006). Despite the fact that the initial
conditions of the particles in the latter model have no relevance to
the invariant manifolds, the maxima of the response spiral arms in
the final state are found to follow closely the underlying pattern
formed by the invariant manifolds.

\begin{acknowledgements}
P. Tsoutsis and C. Kalapotharakos were supported in part by the
Research Committee of the Academy of Athens. We thank an
anonymous referee for the numerous comments which improved the
paper.
\end{acknowledgements}


\noindent{\bf APPENDIX}\\

\noindent{\bf A. Permissible regions for motion}\\

The Hamiltonian (\ref{hamgen}), where $p_r=\dot{r}$ and $p_\theta =
\dot{\theta}+\Omega_p r^2$ defines a curve of zero velocity
$\dot{r}=\dot{\theta}=0$ given by
$$
-{1\over 2}\Omega_p^2r^2+V_0(r)+V_1(r,\theta)=E_J    \eqno{(A1)}
$$
where $V_0(r)$ and $V_1(r,\theta)$ are the total axisymmetric and
non-axisymmetric potential terms respectively. Denoting by $r_s$ the
corotation distance when only axisymmetric potential terms are taken
into account, and $V_s\equiv V_0(r_s)$, we have
$\Omega_p^2=V'_s/r_s$. The curves of zero velocity around $L_4$ and
$L_5$ can be found by expanding the radius and the Jacobi constant
with respect to the corotation values, $r=r_s+(r-r_s)$,
$E_J=-{\Omega^2r_s^2\over 2} + V_s + h$. We find:
$$
2h=-M(r-r_s)^2 +{V_s'''\over 3}(r-r_s)^3+\ldots + 2V_1(r,\theta)
\eqno{(A2)}
$$
where $M=\Omega_p^2-V_0''=4\Omega_p^2-\kappa_s^2$ ($\kappa_s$
denotes the epicyclic frequency).

Consider a very simple form for $V_1$, e.g. $V_1=A\cos(2\theta)$.
For all values of the Jacobi constant between $E_J=E_{J,L_1}$ and
$E_J=E_{J,L_4}$ we have $h-A<0$, thus, since $r-r_s$ is small, we
find from (A2) that $M>0$, or $2\Omega_p/\kappa_s>1$. Omitting terms
of third order in $(r-r_s)$, we find from Eq.(A2)
$$
r-r_s=\pm\bigg\{[2A\cos(2\theta)-2h]/M\bigg\}^{1/2}~~. \eqno{(A3)}
$$
If $-A<h<A$ the motions are outside two banana-like curves around
$L_4$ and $L_5$. In particular, for $r=r_s$ we have, around $L_4$,
$\theta=\pm{1\over 2}\cos^{-1}(h/A)$, while for $\theta=0$ we have
$$
r-r_s\simeq \pm\bigg[2(A-h)/M\bigg]^{1/2}~~.   \eqno{(A4)}
$$
Equation (A4) gives the intersections of the banana-like curves of
zero velocity on the axis connecting the center to the stable
Lagrangian points. The forbidden regions of the CZV can be compared
with the corresponding forbidden regions of the surface of section
$\dot{r}=0$, which is given by Eq.(\ref{discr}). In the latter case
we have $ p_\theta^2=rV'=\Omega^2r^2$, $
\Omega^2r^2-2\Omega\Omega_pr^2+2(V_0+V_1-E_J)=0$. Developing
$r,\Omega$ and $V_0$ in powers of $r-r_s$ and omitting terms of
order higher than the second we find the boundary of the forbidden
region:
$$
\Bigg(-M+{M^2\over
4\Omega_p^2}\Bigg)(r-r_s)^2+2A\cos(2\theta)=2h~~.\eqno{(A5)}
$$
This gives
$$
r-r_s=\pm\bigg[2A\cos(2\theta)-2h\bigg]^{1/2}
{2\Omega_p\over\kappa_s }~~. \eqno{(A6)}
$$
Since $2\Omega_p/\kappa_s>1$,  this boundary is larger than the
boundary (A3).\\

\end{document}